\documentclass[twocolumn]{aastex631}

\usepackage{soul, hyperref}

\usepackage{babel}

\usepackage{tabularx}

\begin{document}

\title{Evaluating Star-Planet Interactions with Zeeman Doppler Imaging: Case Study in YZ~Ceti}

\author[0000-0002-4489-0135]{J. Sebastian Pineda}
\affiliation{University of Colorado Boulder, Laboratory for Atmospheric and Space Physics, 3665 Discovery Drive, Boulder, CO 80303, USA}
\email{sebastian.pineda@lasp.colorado.edu}
\correspondingauthor{J. Sebastian Pineda}

\author[0000-0002-2558-6920]{Stefano Bellotti}
\affiliation{Leiden Observatory, Leiden University,
            PO Box 9513, 2300 RA Leiden, The Netherlands}
\affiliation{Institut de Recherche en Astrophysique et Plan\'etologie,
            Universit\'e de Toulouse, CNRS, IRAP/UMR 5277,
            14 avenue Edouard Belin, F-31400, Toulouse, France}

\author[0000-0003-3924-243X]{Jackie Villadsen}
\affiliation{Bucknell University,
            One Dent Drive, Lewisburg, PA 17837}

\author[0000-0001-5371-2675]{Aline Vidotto}
\affiliation{Leiden Observatory, Leiden University,
            PO Box 9513, 2300 RA Leiden, The Netherlands}

\author[0000-0002-4996-6901]{Julien Morin}
\affiliation{Laboratoire Univers et Particules de Montpellier,
             Universit\'e de Montpellier, CNRS,
             F-34095, Montpellier, France}

\author[0000-0002-9023-7890]{Colin P. Folsom}
\affiliation{Tartu Observatory, University of Tartu, Observatooriumi 1, 61602, Toravere, Estonia}

\begin{abstract}

The recent detections of radio emission from the nearby exoplanet host, YZ~Ceti, suggest that the star is possibly interacting with its rocky innermost planet. These radio emissions are characterized by strong circular polarization, and appear to repeat within consistent orbital phase windows dictated by the orbital position of YZ~Ceti~b. If confirmed, this interaction would provide a first means to concretely assess the magnetic field of a close-in rocky exoplanet. This kind of magnetic star-planet interaction (SPI) should depend on both the exoplanetary orbit, and the geometry of the stellar magnetic field. In this article, we report measurements of the large-scale magnetic field topology of the star YZ~Ceti for the first time, and interpret the cumulative radio data sets in that context to evaluate the plausibility of magnetic SPIs. We find evidence both against and in support of the SPI hypothesis, but crucially that the measured magnetic field does not rule out SPI scenarios. However, clear evaluation of these possibilities requires more accurate assessments of the magnetic field evolution across time. We additionally suggest that YZ~Ceti may be exhibiting planet-induced flaring potentially triggered by exoplanet crossings of the Alfv\'{e}n surface as the planet orbit approaches the stellar magnetic equator, and YZ~Ceti~b experiences dramatic shifts in the ambient field, its polarity, and connectivity to the host star.

\end{abstract}


\section{Introduction} \label{sec:intro}

The recent proliferation of exoplanet detections around nearby low-mass stars has expanded the variety of worlds known and accessible to detailed study and characterization \citep[e.g.,][]{Dressing2015ApJ...807...45D, Ment2023AJ....165..265M}. In particular, a population of close-in planets has emerged which experience dramatically different environments as compared to typical habitable zone worlds. Short-period planets often receive strong photon and particle fluxes with significant consequences for exoplanet atmospheric evolution \citep[e.g., atmospheric mass-loss, photochemistry, etc.; see ][]{Hu2012ApJ...761..166H, Owen2012MNRAS.425.2931O, Rimmer2018SciA....4.3302R,Tilley2019AsBio..19...64T}. Their close-in orbits offer this opportunity to study the different ways that planets are shaped by their stellar environments. 

Notably, magnetic star-planet interactions (SPIs), those mediated by the host-companion magnetic fields, are of great interest because the strength of this interaction depends on the properties of the stellar wind and the magnetic field of the exoplanetary companion \citep[e.g.,][]{Zarka2007P&SS...55..598Z,Saur2013A&A...552A.119S}. Stellar winds have a high degree of uncertainty due to scant observational constraints and significant astrophysical scatter \citep[e.g.,][]{Wood2021ApJ...915...37W,Vidotto2021LRSP...18....3V}, while exoplanet magnetic field measurements remain unconfirmed, especially for rocky planets \citep[see within][]{Brain2024RvMG...90..375B}. The astrophysical potential then from the successful detection and confirmation of magnetic SPIs has driven numerous efforts to search for their corresponding observational signatures \citep[e.g.,][]{Shkolnik2008ApJ...676..628S, Callingham2021NatAs...5.1233C, Narang2023AJ....165....1N, OrtizCeballos2024AJ....168..127O, PenaMonino2024A&A...688A.138P,Villadsen2025_survey}.  

These may include enhanced chromospheric emission and/or flaring modulated at planetary orbital or synodic periods \citep[e.g.,][]{Shkolnik2005ApJ...622.1075S,Cauley2019NatAs...3.1128C, Loyd2023AJ....165..146L, Ilin2024MNRAS.527.3395I}, augmented planetary heating through ohmic dissipation \citep{Strugarek2025A&A...693A.220S}, exoplanet radio aurorae \citep[e.g.,][]{Zarka2007P&SS...55..598Z, Griessmeier2007A&A...475..359G, Vidotto2019MNRAS.488..633V, Turner2021A&A...645A..59T}, or planet-induced radio aurorae from the stellar upper atmosphere \citep[e.g.,][]{Saur2013A&A...552A.119S,Turnpenney2018ApJ...854...72T}. The last of these corresponds to a direct analog of the Jupiter-Io system, and requires sub-Alfv\'{e}nic conditions, i.e., stellar wind and planet properties that enable magnetic field perturbations at the exoplanet position to travel more rapidly than the plasma flow velocity (Alfv\'{e}nic Mach number, $ M_{A} < 1$). Short-period planets are more likely to experience sub-Alfv\'{e}nic conditions and drive strong detectable SPIs.

Recent radio studies have revealed suggestive detections from several stars potentially exhibiting sub-Alfv\'{e}nic magnetic star-planet interactions, including GJ~1151 \citep{Vedantham2020NatAs...4..577V}, Proxima Centauri \citep{PerezTorres2021A&A...645A..77P}, and YZ~Ceti \citep{Pineda2023NatAs...7..569P,Trigilio2023arXiv230500809T}. Nevertheless, firm confirmation of the planet-induced nature of the detected emissions has remained elusive with ongoing follow-up efforts and theory providing greater context to each individual system \citep[e.g.,][]{Kavanagh2021MNRAS.504.1511K,BlancoPozo2023A&A...671A..50B,Narang2024AJ....168..265N}, but without definitive confirmation of planetary modulation of radio detections. 

A challenge in confirming these potential radio emissions from sub-Alfv\'{e}nic SPIs is disentangling planet-induced radio emissions from bursts of purely stellar origin. The highly circularly polarized radio signatures of radio aurorae usually distinguish the electron cyclotron maser (ECM) mechanism from canonical unpolarized gyrosynchrotron stellar flaring \citep[e.g.,][]{Dulk1985ARA&A..23..169D}; however, stellar radio bursts exhibit a wide variety of behavior, including strongly polarized events \citep[e.g.,][]{Zic2020ApJ...905...23Z}, and may be energized by classical flaring, as has been reported on the Sun \citep{Yu2024NatAs...8...50Y}. These concerns are especially problematic around very active low-mass stars \citep[e.g.,][]{Villadsen2019ApJ...871..214V}, but may also apply to slowly rotating planet-hosts as multi-wavelength metrics of activity reveal persistent flaring behavior on even old M-dwarfs \citep[e.g.,][]{France2020AJ....160..237F}. Such older stars may also sometimes exhibit surprisingly strong average surface magnetic field strengths as well \citep{Lehmann2024}. The broad context of radio bursts around low-mass stars across MHz to GHz frequencies needs more attention in order to facilitate interpretation of the ongoing SPI radio searches.

Alternatively, observations of exoplanet host magnetic fields can provide crucial context for evaluating the possibilities of magnetic star-planet interactions. The stellar magnetic field strength and topology serve as important drivers of the stellar environment, dictating the wind properties (particle flux, velocity, etc.), and the total magnetic energy impinging on any exoplanet satellites \citep[e.g.,][]{Vidotto2019MNRAS.488..633V}. These properties set the location of the Alfv\'{e}n surface (where the Mach-Alfven number $M_{A} = 1 $), and determine whether the planets experience sub-Alfv\'{e}nic conditions. The stellar magnetic field strength ($B$) at the source regions sets the radio frequency for potential SPI ECM emissions (as per the electron cyclotron fundamental frequency, $\nu$ [MHz] $\approx 2.8\times B$ [G]), and we would expect SPI radio non-detections while observing far above this frequency. The large-scale field topology provides 3D information about the space environment in relation to planetary orbits, which has been used extensively in studies of exoplanetary space weather \citep[e.g.,][]{Strugarek2015ApJ...815..111S,Garraffo2017ApJ...843L..33G,Elekes2023A&A...671A.133E, Vidotto2023A&A...678A.152V}. This environmental context sets the total amount of energy available for powering magnetic SPIs \citep[e.g.,][]{Zarka2007P&SS...55..598Z,Saur2013A&A...552A.119S}.  For known exoplanets, the field topology enables a clear means of connecting the planetary position to the magnetic footpoint in the stellar corona.

Although the field topology is \textit{a priori} unknown, or often assumed to be predominantly dipolar, recent advances in observing capabilities have enabled new measurements of stellar large-scale fields through Zeeman Doppler Imaging \citep[e.g.,][]{Donati1997MNRAS.291..658D,Morin2010MNRAS.407.2269M,Donati2023MNRAS.525.2015D}. High-resolution spectropolarimetric observations monitoring over stellar rotational timescales can reveal repeatably modulated polarized line profile signatures that depend on the magnetic field strength and the line-of-sight orientation of the magnetic field vector. With careful modeling, these data of stellar disk integrated line profiles can reveal the surface topology and structure of the stellar magnetic field \citep[see within][]{Donati1997MNRAS.291..658D, Lehmann2022MNRAS.514.2333L}. These data sets can thus enable detailed modeling of individual exoplanet systems, their environments, and the corresponding star-planet interactions \citep[e.g.,][]{Kavanagh2021MNRAS.504.1511K}.

With this article, we measure the magnetic field topology of YZ~Ceti, and evaluate the hypothesis of magnetic star-planet interactions between the host and its innermost planet YZ~Ceti~b. \citet{Pineda2023NatAs...7..569P} published the detection of coherent radio bursts from this system noting a recurrence of circularly polarized radio detections within a narrow orbital phase range. Data presented by \citet{Trigilio2023arXiv230500809T} further showed additional radio detections at similar orbital phases. The collective monitoring campaigns showed multiple polarized radio detections in different bands, radio flaring, and extended periods without emission. If some of these bursts are a consequence of magnetic star-planet interactions, then the magnetic field of the system must enable and explain both the detections \emph{and} the non-detections. Conversely, if the topology prohibits the detection of polarized radio emissions in the system, then it may clearly disprove the SPI scenario. 

In Section~\ref{sec:yzcet}, we review the properties of the exoplanet host YZ~Ceti, and present our new spectropolarimetric observations of this target. In Section~\ref{sec:zdi}, we present our analysis of that data, and our assessment of the star's magnetic field topology through Zeeman Doppler Imaging. In Section~\ref{sec:spi}, we use the magnetic field measurement to evaluate the strength of potential magnetic star-planet interactions between YZ~Ceti and its innermost planet. In Section~\ref{sec:geom}, we evaluate the SPI hypothesis in light of the host stellar magnetic field geometry. In Section~\ref{sec:discuss} we discuss our results and finally summarize our findings in Section~\ref{sec:summary}.

\begin{table}[ht]
\begin{center}
\caption{YZ~Ceti System Properties \label{tab:prop}}
\begin{tabular}{l c r}
\hline
Spectral Type & M4.5 & (1)  \\
Distance (pc) & $3.71668$ & (2)  \\
Mass ($M_{\odot}$) & $0.137 \pm 0.003$ & (3) \\
Radius ($R_{\odot}$) & $ 0.163 \pm 0.007$ & (3) \\
$L_{\mathrm{bol}}$ ($10^{30}$ erg s$^{-1}$) & $ 8.6 \pm 0.1 $ & (3)\\
$T_{\mathrm{eff}}$ (K) & $3110 \pm 70 $ & (3)  \\
$\log_{10} \, [L_{\mathrm{X}} / L_{\mathrm{bol}}]$ & $-4.13$ & (4) \\
$\log_{10} \, [L_{\mathrm{H}\alpha}/ L_{\mathrm{bol}}]$ & $-4.32$ & (5) \\
$P_{\mathrm{rot}}$ (d) & $\sim$68.4 & (6) \\
$Bf$ (kG) & $\sim$2.2 & (7) \\
\hline
Planet Periods (d)  & 2.02, 3.06, 4.66 & (6) \\
Planet $a/R_{*}$& 21.6, 28.4, 37.6  & (6) \\
Planet $m_{p} \sin i_{orb}$ ($M_{\oplus}$)  & 0.70, 1.14, 1.09 & (6) \\
\hline 
\multicolumn{3}{p{0.9\linewidth}}{ R\textsc{eferences}: (1) \citet{Reid1995}; (2) \citet{Gaia2020yCat.1350....0G}; (3) \citet{Pineda2023NatAs...7..569P}; (4) \citet{Stelzer2013}; (5) \citet{Reiners2018}; (6) \citet{Stock2020}; (7) \citet{Moutou2017} }\\

\end{tabular}
\end{center}
\end{table}

\section{Target: YZ Ceti} \label{sec:yzcet}

YZ~Ceti is a very near ($<$4~pc) fully convective late M-dwarf star that is slowly rotating (see Table~\ref{tab:prop}). Its mass and rotation period place it in the unsaturated portion of the rotation-activity correlation, with moderate to weak chromospheric and coronal emissions as traced by H$\alpha$ and X-rays \citep[e.g., see within][]{Newton2017ApJ...834...85N,Wright2018MNRAS.479.2351W}. The literature measurement of the average surface magnetic field yielded a value of 2.2~kG from Zeeman broadening \citep{Moutou2017}, but that measurement may be systematically high \citep{Reiners2022A&A...662A..41R}, although some of these stars at slow rotation periods likely possess stronger fields than anticipated \citep{Lehmann2024}.

Radial velocity monitoring of the star revealed the presence of at least 3 planets in the system, likely near or greater than Earth in mass \citep{AstudilloDefru2017A&A...605L..11A, Stock2020}. Furthermore, the planets do not appear to transit the star, limiting estimates of the exoplanet sizes \citep{Stock2020}. This analysis yields a maximum orbital inclination of the planets as $i_{orb} < 87.43^{\circ}$. Based on the orbital solutions, the innermost planet YZ~Ceti~b passes inferior conjunction (when it would transit if orbit inclination was larger) at $t_{b} = 2452996.25$ (BJD).

The polarized radio detections of the system showed a recurrence of radio bursts at similar orbital phases typically within a quarter period after inferior conjunction. The first detections, made with the Karl G.~Jansky Very Large Array (VLA) in the 2-4 GHz band, took place between 30 November, 2019 and 29 February, 2020 \citep{Pineda2023NatAs...7..569P}. Additional radio detections made with the uGMRT at 550-900 MHz, took place between 1 May, 2022 and 3 September, 2022 \citep{Trigilio2023arXiv230500809T}. We examine the geometry during these radio epochs in conjunction with the host magnetic field topology in Section~\ref{sec:geom}.

\subsection{SPIRou Data}\label{sec:spirou}

For the magnetic field characterization, we analyzed spectropolarimetric observations collected with the SpectroPolarim\`etre InfraRouge \citep[SPIRou;][]{Donati2020} in circular polarization mode. SPIRou is the stabilized high-resolution near-infrared spectropolarimeter mounted on the 3.6\,m Canada–France–Hawaii Telescope (CFHT) atop Maunakea, Hawaii. SPIRou operates between 0.96 and 2.5~$\mu$m ($YJHK$ bands) at a spectral resolving power of $R \sim 70\,000 $. A polarimetric sequence is obtained from four consecutive sub-exposures. Each sub-exposure is taken with a different rotation of the retarder waveplate of the polarimeter relative to the optical axis. In circular polarization mode, the output is given by the total intensity spectrum (Stokes~$I$) and the circularly polarized spectrum (Stokes~$V$). Optimal extraction of Stokes~$I$ and Stokes~$V$ spectra was carried out with {\it A PipelinE to Reduce Observations} (\texttt{APERO}) version 0.7.288, a fully automatic reduction package installed at CFHT \citep{Cook2022}. YZ~Ceti was visited 43 times between October 2023 and January 2024, spanning 90 days in total. Considering that the H band magnitude of YZ~Ceti is 6.75, a total exposure time of 600\,s per observation was used, corresponding to four polarimetric sub-exposures of 150\,s. This yielded a signal-to-noise ratio (S/N) at 1650~nm per spectral element between 143 and 295 for our observations, with a mean value of 260. The journal of the observations is given in the Appendix (Table~\ref{tab:spirou_log}).

\begin{figure*}[ht]
    \centering
    \includegraphics[width=\textwidth]{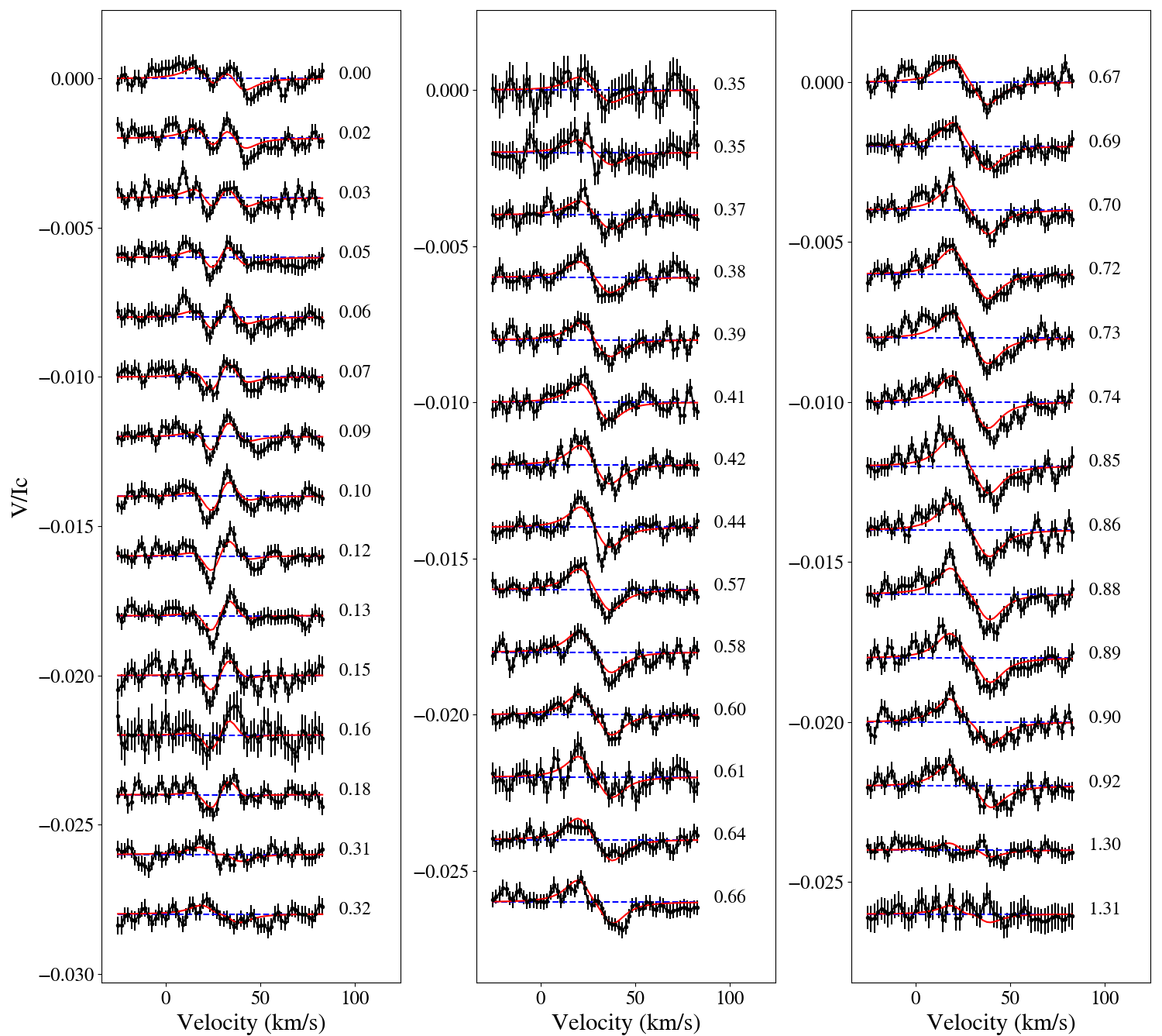}
    \caption{Time series of Stokes~$V$ LSD profiles of YZ~Ceti. Black clurves indicate the observations and red lines represent the ZDI models for P$_\mathrm{rot}=68.4$\,d, $v_\mathrm{eq}\sin(i)=0.1$\,km\,s$^{-1}$, $i=60^\circ$, and $\mathrm{d}\Omega=0.0$\,rad\,d$^{-1}$. The horizontal line represents the zero point of the profiles, which are shifted vertically based on their rotational phase for visualization purposes.}
    \label{fig:stokesV}
\end{figure*}

\section{Zeeman Doppler Imaging}\label{sec:zdi}

\subsection{Model and input parameters}

To detect and characterize polarization signatures induced by the Zeeman effect, we applied least-squares deconvolution \citep[LSD;][]{Donati1997MNRAS.291..658D, Kochukhov2010a} to unpolarised and circularly polarized spectra using the \textsc{lsdpy} code which is part of the Specpolflow software \citep{Folsom2025}\footnote{The python LSD code is available at \href{https://github.com/folsomcp/LSDpy}{https://github.com/folsomcp/LSDpy} (version 1.0.0)}. This numerical technique deconvolves the observed spectrum with a line list and outputs an pseudo-average line profile with substantially reduced noise. We used a line list similar to \citet{Bellotti2024a}, but additionally restricting the selection to the 105 atomic lines with the least blending from molecular and telluric features, and without very strong wings from pressure broadening, in order to minimize distortions in the LSD profile. We applied a normalizing wavelength of 1650\,nm and Land\'e factor of 1.2. The line list corresponds to a local thermodynamic equilibrium model \citep{Gustafsson2008} with $T_{\mathrm{eff}}=3000$\,K, $\log g=5.0$\,[cm\,s$^{-2}$], $v_{\mathrm{micro}}=1$\,km\,s$^{-1}$, and contains lines between 950 nm and 2600 nm. The line list was synthesized using the Vienna Atomic Line Database\footnote{\url{http://vald.astro.uu.se/} using the Montpellier mirror to locally request MARCS model atmospheres.} \citep[VALD,][]{Ryabchikova2015}. We obtained magnetic detections in all but three observations, where the non-detections are due to stellar rotation and large-scale magnetic field geometry as outlined below. The magnetic field reconstruction is performed considering a range of $\pm55$\,km\,s$^{-1}$ with respect to the systemic radial velocity center of the Stokes~$I$ LSD profiles at 28.3\,km\,s$^{-1}$, which also encompasses the magnetic signature present in the Stokes~$V$ LSD profiles.

\begin{figure*}[htb]
    \centering
    \includegraphics[width=\textwidth]{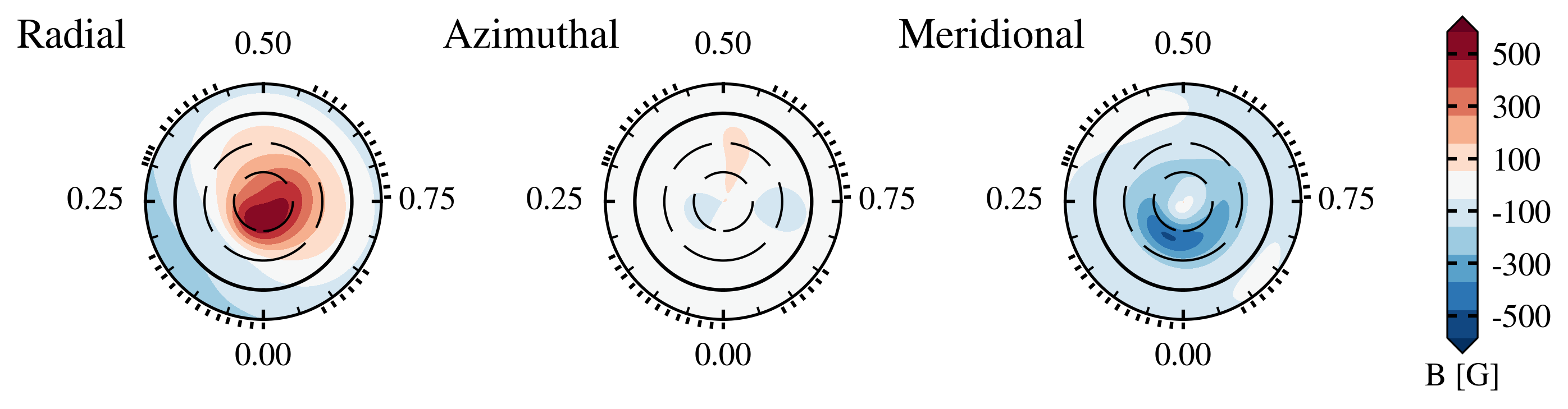}
    \caption{Reconstructed ZDI map of the large-scale magnetic field of YZ~Ceti. From the left, the radial, azimuthal, and meridional components of the magnetic field vector are illustrated. Concentric circles represent different stellar latitudes: -30\,$^{\circ}$, +30\,$^{\circ}$, and +60\,$^{\circ}$ (dashed lines), as well as the equator (solid line). The radial ticks are located at the rotational phases when the observations were collected. The color bar indicates the polarity and strength (in G) of the magnetic field.}
    \label{fig:zdi}
\end{figure*}

We characterized the large-scale magnetic field strength and topology of YZ~Ceti with Zeeman-Doppler imaging \citep[ZDI;][]{Semel1989,DonatiBrown1997}, using the python implementation \texttt{zdipy}\footnote{Available on Github \href{https://github.com/folsomcp/ZDIpy}{https://github.com/folsomcp/ZDIpy}} with the inclusion of the Unno-Rachkovsky's solutions to polarized radiative transfer equations \citep[see][]{Folsom2018,Bellotti2024a} and the filling factor formalism of \citet{Morin2008}. The magnetic field vector is modeled as the sum of a poloidal and a toroidal component, which are both expressed with spherical harmonics formalism \citep{Donati2006MNRAS.370..629D,Lehmann2022MNRAS.514.2333L}. The ZDI technique adopts a fitting routine that iteratively models the rotationally modulated Stokes~$V$ profiles, until a selected reduced $\chi^2$ is achieved. Because the inversion problem is ill-posed, a maximum-entropy regularization scheme is applied to obtain the field map compatible with the data and with the lowest information content \citep[for more details see][]{Skilling1984,DonatiBrown1997}. 

The intrinsic Stokes~$I$ profile shape is fitted with the Unno-Rachkovsky model \citep[see e.g.][]{delToroIniesta2003,Landi2004}, described by a Gaussian width of 0.1\,km\,s$^{-1}$, a Lorentzian width of 6.6\,km\,s$^{-1}$, and a line-to-continuum absorption coefficients ratio of 15.2. These values were obtained by $\chi^2$ minimization applied on the median Stokes~$I$ LSD profile. We note that the Gaussian broadening is unresolved given the instrumental resolution and the other line broadening processes (in particular, Lorentzian and Zeeman broadening). Testing a value of 1.0\,km\,s$^{-1}$ for the Gaussian width produced identical results, confirming the lack of a strong constraint. The slope of the source function in the Milne-Eddington atmosphere was set to 0.25 \citep[see][]{Bellotti2024a}, consistently with the adopted linear limb darkening coefficient in H band of 0.2 \citep{Claret2011}. The rotation period of YZ~Ceti has been estimated to be approximately 68.4\,d from photometry and spectroscopic stellar activity indicators \citep{Stock2020,Lafarga2021}. The projected equatorial velocity $v_\mathrm{eq}\sin(i)$ is less constrained, with an upper limit around 2\,km\,s$^{-1}$ \citep{Reiners2018,Stock2020}. From geometrical considerations, the combination of the rotation period and the stellar radius $0.163\pm0.007$\,R$_\odot$ results in $v_\mathrm{eq}\sin(i)\simeq0.1$\,km\,s$^{-1}$. We set the maximum degree of spherical harmonic coefficients $\ell_\mathrm{max}$ to 10, although most of the magnetic energy is contained in the $\ell\leq4$ modes.

Assuming $v_\mathrm{eq}\sin(i)=0.1$\,km\,s$^{-1}$ and fixing the other parameters, we used ZDI to constrain the stellar inclination, by generating maps over a grid of inclinations and selecting the value minimizing the $\chi^2$. We found an inclination of $60\pm5^\circ$, which is in agreement with the upper limit estimated by \citet{Reiners2018}. Then, following the method of \citet{Petit2002}, the search for latitudinal differential rotation was inconclusive, hence we assumed a single rotation rate for the stellar surface in the ZDI reconstructions. Finally, we attempted a search of filling factors on Stokes~I ($f_I$) and on Stokes~$V$ ($f_V$) similarly to \citet{Bellotti2024a}, but without success. We then decided to fix $f_I=0.5$ and $f_V=0.15$ assuming typical values that have been constrained for M~dwarfs of similar spectral type as YZ~Cet \citep[see e.g.][]{Morin2008,Morin2010MNRAS.407.2269M,Donati2023,Bellotti2024a}.

\subsection{Reconstructed topology}

The Stokes~$V$ model profiles were fit to a $\chi^2_r$ of 0.8 from an initial value of 1.9, as shown in Figure~\ref{fig:stokesV}. The fact that the $\chi^2_r$ is lower than one stems most likely by overestimated error bars in the uncertainty propagation of the flux applied by the pipeline. The ZDI map of the large-scale magnetic field of YZ~Ceti is shown in Figure~\ref{fig:zdi}. The topology features a predominantly poloidal configuration, storing 99.0\% of the magnetic energy. This is consistent with the fact that, at very low $v_\mathrm{eq}\sin(i)$, we are weakly sensitive to the large-scale toroidal field component \citep[as discussed also in][]{Lehmann2022MNRAS.514.2333L}. This has only a small effect on the study in this article, as the field extrapolated throughout the exoplanetary system is dominated by the radial field component. The very low $v_\mathrm{eq}\sin(i)$ also means that spherical harmonics above $\ell = 3$ are largely unresolved \citep{Morin2010MNRAS.407.2269M}, and these missing components may have an impact on the extrapolated field near the surface of the star. Of the poloidal component, 71\% lies in the dipolar mode, 15\% in the quadrupolar mode, and 8\% in the octupolar mode. In addition, the field stores 87\% of the energy in the axisymmetric modes (that is $\ell\geq1$, $m=0$). The average magnetic field strength is 225\,G (peak field of 560~G in positive north pole, see Figure~\ref{fig:zdi}), which is in agreement with the recent trend discovered within the SPIRou Legacy Survey \citep{Lehmann2024}, for which slowly rotating (around 100\,d rotation period) M~dwarfs can sustain intense magnetic fields on the order of hundreds of Gauss. The topology is mostly dipolar, but also shows a strong north-south asymmetry, with a southern pole weaker by a factor of $\sim$2. This, however, may be partially an artifact of the maximum-entropy regularization scheme in the ZDI technique: the lowest energy solution prefers a weak southern pole, where we lack concrete observational constraints due to the viewing geometry. 

When compared to other stars with similar masses and rotation periods, and published ZDI maps, YZ~Ceti is similar to Proxima Cen and Gl~905 \citep{Klein2021,Lehmann2024}. Proxima Cen is a M5.5 star with a mass of 0.12\,M$_\odot$ and $P_\mathrm{rot}=89.8$\,d. The ZDI reconstruction by \citet{Klein2021} exhibited a mostly poloidal, dipolar, and non-axisymmetric field with an average strength of 200\,G. Besides the axisymmetry, these characteristics are similar to our large-scale field reconstruction of YZ~Ceti. Gl~905 is an M5.0 star with a mass of 0.15\,M$_\odot$ and $P_\mathrm{rot}=114.3$\,d. \citet{Lehmann2024} performed three ZDI reconstructions yearly between 2019 and 2022, showing that the predominantly poloidal-dipolar magnetic field configuration featured a substantial reduction in axisymmetry from 70\% to 4\%, and the average field strength also decreased from 130\,G to 60\,G. Significant field evolution between epochs could present challenges for interpretation of star-planet interactions as the radio data sets occur before our ZDI measurements (see Section~\ref{sec:geom}). We do not yet know how YZ~Ceti's field evolved overtime; we discuss these issues in Section~\ref{sec:discuss}.

\section{Revising SPI Radio Predictions} \label{sec:spi}

\subsection{Field Extrapolation and Radio Flux}\label{sec:pfss}

Our new magnetic field topology enables a first test of the SPI scenario: is the predicted interaction strength consistent with the measured radio flux densities? In this article, we focus the SPI scenario on YZ~Ceti~b, and address considerations for the outer two planets in Appendix~\ref{sec:ap}. 

\begin{figure}[tb]
    \centering
    \includegraphics[width=0.5\textwidth]{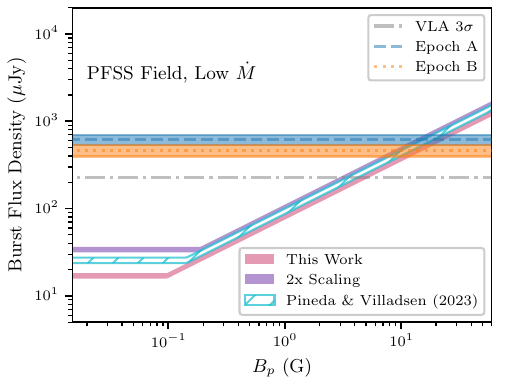}
    \caption{Predicted strengths for planet-induced radio bursts from YZ~Ceti require a potentially strong planetary field for YZ~Ceti~b to match detections with VLA (Epochs A and B). The three curves compare the previous result from \citet{Pineda2023NatAs...7..569P} to this work utilizing the new magnetic field measurements of YZ~Ceti. The top trend (purple) further examines the implied predictions if the measured radial magnetic field were scaled by a factor of 2. VLA sensitivity is shown for 3~min integrations. }
    \label{fig:predbplanet}
\end{figure}

In \citet{Pineda2023NatAs...7..569P}, we used a scaled version of Proxima Centauri's magnetic field as a proxy for what we could expect from YZ~Ceti. We update that analysis now based on the newly measured field strengths (see Section~\ref{sec:zdi}), and use the full 3D topology in Section~\ref{sec:geom} for assessments of SPI. Our previous assumption for the surface average total field strength was $\sim$280~G, and now the map implies a value of $\sim$225~G. As in \citet{Pineda2023NatAs...7..569P}, this analysis collapses the magnetic field information into an average estimate of the exoplanetary wind environment to quickly enable calculations of the strength of star-planet interactions. We summarize the process for predicting the flux density of planet-induced polarized radio bursts as follows. First, we used the ZDI information to extrapolate the surface magnetic field out to the stellar corona, assuming a potential field source surface model \citep[PFSS; e.g.,][]{Altschuler1969SoPh....9..131A,See2017MNRAS.466.1542S}, mimicking the closed topology of the stellar corona out to a distance of $R_{s} = 5R_{*}$. Second, we used the average magnetic field strength at the source surface ($R_{s}$) as the magnetic field boundary condition to numerically solve a Weber-Davis isothermal wind \citep{Weber1967ApJ...148..217W,Pineda2023NatAs...7..569P}. The wind solution defines the properties of the magnetized environment throughout the exoplanet system. Lastly, we consider the energy dissipated across the planetary magnetospheric obstacle using an Alfv\'{e}n wing prescription \citep{Saur2013A&A...552A.119S}, and convert the available energy to an expected radio burst flux density, see Figure~\ref{fig:predbplanet}. 

The potential field source surface extrapolation allows for a realistic 3D topology anchored in the measured radial field, but assumes that the field is minimally strained by the coronal plasma motions. While an approximation, this assumed topology captures the crucial close-in decay of the stellar magnetic field strength (e.g., $B \propto r^{-3}$ for a dipolar field), which is the key feature needed in assessing the stellar magnetic field at the location of the exoplanets. Beyond the source surface, $R_{s}$,  the magnetic field decays as $B \propto r^{-2}$. The PFSS formalism has been used readily for quick assessments on the influence of the stellar field topology on X-ray coronae, angular momentum evolution, and mass-loss rates \citep[e.g.,][]{Jardine2002MNRAS.336.1364J,See2017MNRAS.466.1542S}, and for studying the Solar corona \textit{in situ} \citep[e.g.,][]{Panasenco2020ApJS..246...54P}. The source surface, where the field is taken to be effectively radial, is the typical transition point between a fully closed topology and the open stellar wind flow. While unknown, values of $\sim$3-5$R_{*}$ are consistent with treatments of the Sun, and other low-mass stars using MHD simulations \citep{Vidotto2014MNRAS.438.1162V, See2017MNRAS.466.1542S}. Our assumption of $R_{s} = 5R_{*}$ is based on the expectation that dwarfs with stronger field strengths are likely to exhibit a larger volume of closed magnetic field. The stellar magnetic field strength at the planet location depends inversely on the assumed $R_{s}$, but should not deviate significantly from our conservative choice, with differences potentially on the order of 20\%.

The Weber-Davis equatorial wind solution is similar to a Parker-style stellar wind \citep{Parker1958ApJ...128..664P}, but self-consistently solves the coupling of the magnetic field to the outflowing plasma in an equatorial azimuthally symmetric geometry. We focused on the weaker wind case for this slowly rotating star, based on observed M-dwarf properties \citep[see within][]{Pineda2023NatAs...7..569P},  utilizing a coronal temperature of $kT = 0.25$~keV, and a mass-loss rate of 0.25$\dot{M}_{\odot} = 5\times10^{-15} M_{\odot}$ yr$^{-1}$. The updates to the input magnetic field have little impact on the assumed radial wind profile, as expected,  yielding similar wind velocities (600-700 km s$^{-1}$) at the location of the planets compared to past results \citep[see supplementary information in ][]{Pineda2023NatAs...7..569P}. Updates to the SPI calculation are driven by the changes in the magnetic field strength at the planet orbit.

With well defined magnetized wind properties, as in \citet{Pineda2023NatAs...7..569P}, we determined the amount of energy transmitted back to the host star via star-planet interactions following \citet{Saur2013A&A...552A.119S} in cgs units:

\begin{equation}
    S = \bar{\alpha}^{2} R_{o}^{2} \upsilon^{2} B \sin^{2} \theta \sqrt{\pi \rho} \; ,
    \label{eq:saur}
\end{equation}

\noindent where $R_{o}$ is the effective size of the planetary obstacle comparable to the magnetopause standoff distance determined through pressure balance with the stellar wind, $\upsilon$ is the plasma flow speed in the planet frame, $B$ is the stellar magnetic field strength at the exoplanet location, $\rho$ is the mass density of the plasma, $\theta$ is the angle between the flow speed and the magnetic field, and $\bar{\alpha}$ is an interaction strength parameter that goes to unity for magnetized planets and/or those with strong ionospheres \citep[see also][]{Turnpenney2018ApJ...854...72T}. For intrinsically weak planetary magnetic fields relative to the wind field strength, the obstacle size reduces to the exoplanet radius.

We express the obstacle size explicitly as a function of the planetary dipole field, $B_{p}$, and the wind properties as

\begin{equation}
    R_{o} = k_{w} \, R_{p} \left(  \frac{  B_{p}^{2}  }{ 8\pi \rho \upsilon^{2}  + 8\pi \rho k T/\mu m_p  + B^{2}  }      \right)^{1/6} \; ,
    \label{eq:ro}
\end{equation}

\noindent where, $R_p$  is the planet radius, $m_p$ is the proton mass, $\mu = 0.5$ for the fully ionized wind, and we utilized the geometric factor of $k_{w} = \sqrt{3}$ for the effective radius of the Alfv\'{e}n wings from the interaction of the wind with the planetary magnetosphere \citep{Saur2013A&A...552A.119S}.

We converted the available energy to a radio flux density applying typical literature prescriptions:

\begin{equation}
    F_{\nu} = \frac{ \epsilon S}{\Omega \Delta \nu d^{2}} \; ,
    \label{eq:rflux}
\end{equation}

\noindent where $S$ comes from Equation~\ref{eq:saur}, $\epsilon = 0.01$ is efficiency factor of energy conversion to ECM radio emission  \citep[e.g.,][]{Zarka2007P&SS...55..598Z,Saur2013A&A...552A.119S, Callingham2021NatAs...5.1233C}, $\Delta \nu = 3$~GHz is the emission bandwith, consistent with VLA data sets, $d = 3.71668$~pc is the distance to the star, and we use $\Omega = 0.16$~sr for the beaming angle based on Jupiter-Io radio observations \citep{Queinnec2001P&SS...49..365Q}. 

The results of our calculations are displayed in Figure~\ref{fig:predbplanet} as a function of the assumed strength of YZ~Ceti~b's intrinsic planetary dipolar field. The width of the shaded results (purple/red) corresponds to changes in the planet radius from 1~$R_{\oplus}$ down to the radius matching an Earth-like density at the observed minimum mass (see Table~\ref{tab:prop}), while taking the orbital inclination as matching that of the rotational inclination ($i = 60^{\circ}$, see Section~\ref{sec:zdi}). The bottom-most curve (red) utilizes the updated field strengths. For YZ~Ceti~b the magnetopause radius ranges from 2.2-4.7 planetary radii, across dipolar fields of 1-10~G. Relative to the scaled magnetic field topology from Proxima Centauri (blue hatched), the new measurements give slightly weaker stellar wind field strengths throughout the exoplanetary system. The result translates to a weaker prediction for the radio flux, or equivalently a higher requirement for the planetary field, $\sim$15-22~G, to achieve radio bursts matching the observed flux densities.

\begin{figure*}[ht]
    \centering
    \includegraphics[width=0.46\textwidth]{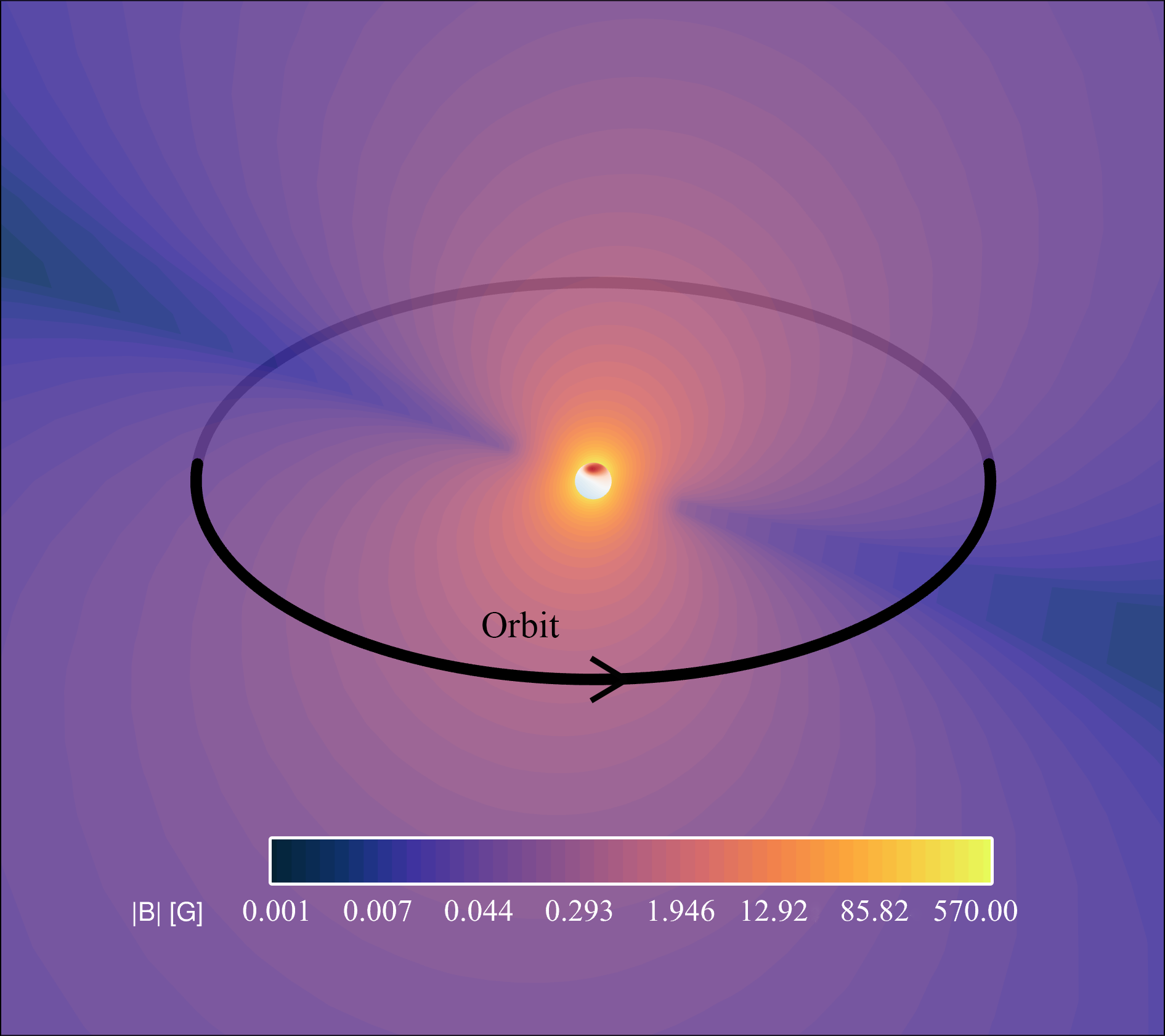}   \includegraphics[width=0.46\textwidth]{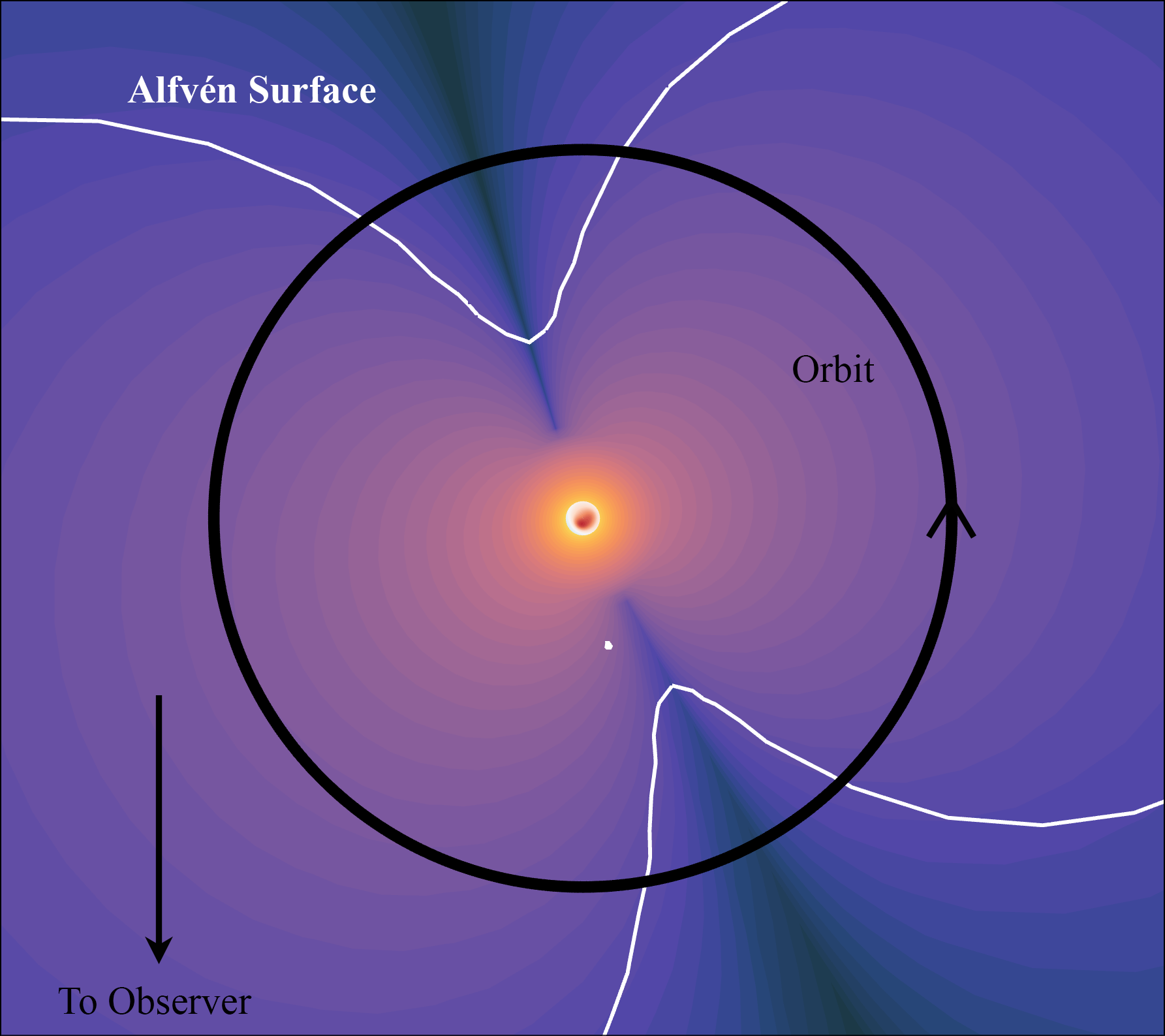}
    \caption{Illustrations of 3D prograde orbit for YZ~Ceti~b aligned with stellar rotational axis, showing strength of magnetic field in plane of sky (left) and in plane of orbit (right); the same color bar for the field is used in both panels. \textit{Left} - system orientation as viewed from Earth. When the planet is in front of the star (darker black trace) the magnetic field lines connecting the exoplanet and the star can trace to visible surface footpoints. \textit{Right} - system viewed from above, perpendicular to orbit plane, with the location of the Alfv\'{e}n surface indicated in white. The orbit is typically in the sub-Alfv\'{e}nic regime (within white contour), but may intersect the Alfv\'{e}n surface. Along the magnetic equator, close to the wind current sheet where the radial field switches polarity, the weakening magnetic field pushes the Alfv\'{e}n surface inward cutting a wedge across the planet orbit. The location of this equatorial wedges rotates with the stellar magnetic field, here shown at the initial ZDI epoch.}
    \label{fig:orborient}
\end{figure*}

\subsection{Scaling the Field Strength}\label{sec:scaleBspi}

The observed maximum magnetic field strength inferred from ZDI for YZ~Ceti was 560~G (see Section~\ref{sec:zdi}). The corresponding cyclotron frequency for the ECM fundamental is $\sim$1.57~GHz. This field would be too weak to explain the VLA detections which extend up to $\sim$3~GHz \citep{Pineda2023NatAs...7..569P}. However, 1.57~GHz exceeds the top of the spectral band corresponding to the uGMRT detections, which show spectra that likely extend beyond their limit at 900~MHz \citep{Trigilio2023arXiv230500809T}. While the electron cyclotron maser instability may produce emission at higher harmonics (multiples of $\sim$2-3 of the fundamental), these modes are typically less efficient \citep{Treumann2006A&ARv..13..229T}.

Because field cancellation in spectropolarimetry \citep[e.g.,][]{Sinjan2024A&A...690A.341S} can miss small scale magnetic structures, it is not necessarily surprising that surface fields could be higher than observed with ZDI. But, if the radio emission is powered by star-planet interaction, then it does imply that electrons accelerated along the magnetic field connecting the star to the planet map to emission regions associated with complex near surface structure of higher field strengths. The details of such a topology are uncertain. Alternative mechanisms, however, could explain the apparent discrepancy in the observed field as compared to the ECM emission frequencies.

 Irrespective of small-scale field cancellation, the observed Stokes~$V$ profiles may still under predict the line-of-sight magnetic field \citep{Rosen2015ApJ...805..169R, Lehmann2019MNRAS.483.5246L}. Moreover, for YZ~Ceti, the viewing geometry and the ZDI solution may be suppressing the strength of the field in the southern magnetic pole, which also biases down the strength of the overall surface averaged field (see Section~\ref{sec:zdi}). Observational biases such as these, up to factors of $\sim$2, are likely impacting all ZDI results. To account for these possible effects in the expected radio flux predictions, we further considered a case where we scaled up the observed field by a factor of 2 so as to match the peak field in the ZDI map to the maximum observed frequency in the radio data. This result is shown with the purple region in Figure~\ref{fig:predbplanet}. Using this scaled stellar field, the dipole magnetic field strength of YZ~Ceti~b would have to be $\sim$8-14~G to generate radio bursts matching the observed flux densities. While still strong, this lower field moves closer towards Earth's 0.5~G field. Exoplanet magnetic fields remain highly uncertain \citep[e.g.,][]{Brain2024RvMG...90..375B}; such constraints will be important for guiding the application of dynamo theory across the exoplanet variety exhibited by the known population.

Our spectropolarimetric data sets were also acquired 2-4 years after the radio epochs. While the new data set does give insight into the expected magnetic field topology, we should also expect field evolution over that time based on monitoring data from similar stars. These changes include overall field strength (factors of a couple), and even changes to the topology degree of axisymmetry \citep{Lehmann2024}. The long-term evolution of YZ~Ceti is unclear and will require additional spectropolarimetric epochs to assess.

\section{Geometry of Star-Planet Interaction} \label{sec:geom}

In addition to constraining plausible strengths for planet-induced radio bursts, direct host magnetic field topology measurements enable a close examination of the geometry of star-planet interaction, and the detectability of anisotropic polarized radio emissions. 

Electron cyclotron maser emission is beamed into a hollow conical region with a large opening angle such that the emission is focused near perpendicular to the direction of the magnetic field vector at the source location \citep{MelroeDulk1982ApJ...259..844M}. The angle depends on the energy of the electron beams, and the width of the cone is thin (1-2$^{\circ}$). Potential planet-induced ECM source regions are expected to be near the stellar surface and located on magnetic field lines connecting the star to the planet, YZ~Ceti~b. In this section, we combine the full topology information (including modes through $\ell=10$), with the radio epochs to examine the properties of the magnetic footpoints connecting YZ~Ceti~b to its host star.

\begin{figure*}
    \centering
    \includegraphics[]{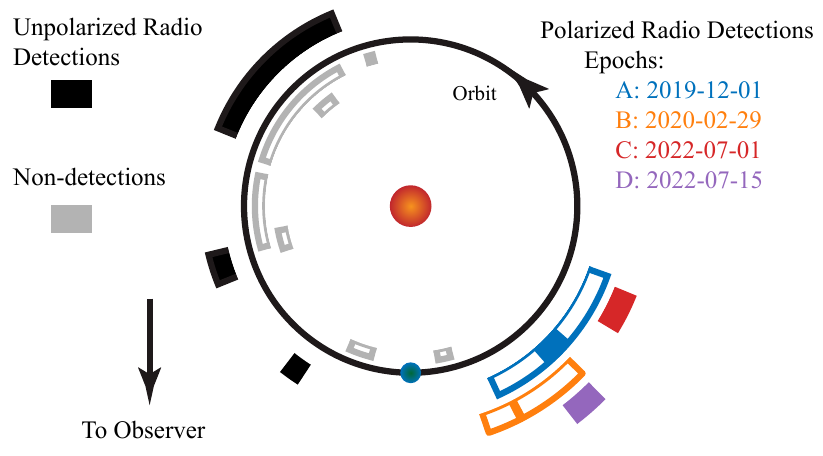}
    \caption{A diagram of the radio epochs (see Table~\ref{tab:radio}), with durations arranged as arcs according to the orbital phase of the exoplanet  YZ~Ceti~b. Longer arcs correspond to \citet{Pineda2023NatAs...7..569P} and shorter arcs to \citet{Trigilio2023arXiv230500809T}. The blue dot shows the planet position at inferior conjunction. The circularly polarized radio detections are clustered in phase just in the lower right, denoted by different colors. Each arc shows both the monitoring length of those epochs, with burst durations as filled arcs (e.g., blue/orange subarcs). Unpolarized detections are shaded black, with non-detections shown in gray. }
    \label{fig:orbdiag}
\end{figure*}

\subsection{Radio Epochs and the Star-Planet Phasing}

To consider how the stellar magnetic field connects the host to the exoplanet at different times, we need to know where the planet is in 3D space at a given epoch, and the relative phasing of the rotating star. The exoplanet orbit solution for YZ~Ceti~b defines the former \citep{AstudilloDefru2017A&A...605L..11A,Stock2020}, and we can use our ZDI topology to define the corresponding stellar rotational phase (see Section~\ref{sec:zdi}), knowing the rotation period of $68.4 \pm 0.05$~d \citep{Stock2020}.\footnote{\citet{Stock2020} report multiple periods, we proceed with their blue-optical result, the most consistent measurements and those expected to be most sensitive to just the contrasts induced by rotating spots.}

The orbital solution for YZ~Ceti~b reveals an exoplanet with a likely circular orbit, that does not transit the star. For this RV detected planet, we have an unknown orientation for the 3D orbit of the system. As a first consideration, we took the orbital angular momentum vector to be aligned with the stellar rotation axis. Based on the best ZDI solution, we assumed an orbital inclination of 60$^{\circ}$ to match the rotational tilt with a prograde orbit. Most exoplanetary systems, especially close-in small planets, although not perfectly aligned, do show agreement between the host rotational axis and the system orbital angular momentum \citep[e.g.,][]{Albrecht2013ApJ...771...11A, Rice2023AJ....166..266R}. With this alignment, and the rotation axis pointing `up', we can visualize the orbit as being counter-clockwise with the planet being `below' the stellar position at the time of inferior conjunction, see left panel of Figure~\ref{fig:orborient}.

We further verified that YZ~Ceti~b experiences sub-Alfv\'{e}nic conditions with this orbit. Using the magnetic field extrapolation in 3D space throughout the exoplanetary system and the expected wind properties of the star (see Section~\ref{sec:spi}), we estimated the location of the Alfv\'{e}n surface exterior to which the planet loses connectivity with its host. We illustrate this feature in the right panel of Figure~\ref{fig:orborient} with a white line in the planetary orbital plane. YZ~Ceti~b should be magnetically connected with its host for most of the orbital period except for when the orbit approaches the magnetic equator and the wind current sheet. Here the stellar magnetic field strength drops as the radial field switches polarity. Because the wind and magnetic field are defined independently here, the exact location of the Alfv\'{e}n surface is approximate; however, it does provide a useful guide for the kinds of interactions we may expect. A full simulation of the 3D environment and wind using the magnetic topology will be required to provide a more detailed assessment. These regions of intersection rotate around with the stellar period, and we discuss their role in SPI further in Section~\ref{sec:discuss}.

The absolute position uncertainty from the orbital ephemeris is $\sim$5~hr (10\% in orbital phase, $\phi_{b}$). At the VLA epochs, relative to the absolution position, there is an additional $\sim$20~min ($\Delta \phi_{b}$$\sim$0.007) uncertainty from the precision in the period (we use the stable 3-planet solution from \citet{Stock2020}: $P=2.02087 \pm^{0:00007}_{0:00009}$, $t_{b} = 2452996.25\pm^{0.21}_{0.17}$~BJD). This grows to $\sim$80~min ($\Delta \phi_{b}$$\lesssim$0.03) at the uGMRT epochs. If we phase the star backwards to the radio epochs it introduces a rotational phase uncertainty in the ZDI topology of $\sim$0.015 and $\sim$0.005, at the VLA and uGMRT epochs respectively. 

Given those caveats, we visualize the aggregated radio observations in Figure~\ref{fig:orbdiag}. Around the circle representing the orbit of YZ~Ceti~b, we show arcs corresponding to each of the 14 radio observations indicating their phasing relative to the time of inferior conjunction, with the arc length corresponding to the duration of that data set. In Table~\ref{tab:radio} we show the labeled dates and polarization properties of the relevant radio data discussed in this article. In Figure~\ref{fig:orbdiag}, we show all detections on the exterior (black/color), and all non-detection (gray) interior to the circle, while distinguishing polarized (color) from unpolarized (black) events. The four polarized burst detections are all clustered in the bottom right quadrant. Moreover, events A and D overlap at nearly the same exact orbital phase of YZ~Ceti~b. 

While suggestive, for these events to be planet-induced they should also be consistent with the magnetic field topology, and that topology should further help explain the absence of polarized radio bursts at all other epochs.

\newcolumntype{L}[1]{>{\hsize=#1\hsize\raggedright\arraybackslash}X}%
\newcolumntype{R}[1]{>{\hsize=#1\hsize\raggedleft\arraybackslash}X}%
\newcolumntype{C}[1]{>{\hsize=#1\hsize\centering\arraybackslash}X}%

\begin{table}[ht]
\begin{center}
\caption{Summary of Literature Radio Epochs \label{tab:radio}}
\begin{tabularx}{\linewidth}{ L{1}  C{1}  C{1} c }
\hline
\hline
\multicolumn{4}{p{\linewidth}}{ \centering VLA \citep{Pineda2023NatAs...7..569P} }\\
\hline
Date & Detection & Polarization$^{*}$ & PV\# \\
\hline
2019-11-30 & $\times$ &  --- & 1\\
2019-12-01 & \checkmark & R, U$^{\dagger}$ & 2/A \\
2019-12-02 & \checkmark & U & 3 \\
2020-02-02 & $\times$ & --- & 4 \\
2020-02-29 & \checkmark & L $^{\ddagger}$ & 5/B \\
\hline
\hline\\
\multicolumn{4}{p{\linewidth}}{ \centering uGMRT \citep{Trigilio2023arXiv230500809T} }\\
\hline
Date & Detection & Polarization & T\# \\
\hline
2022-05-01 & $\times$ & --- & 1\\
2022-05-16 & $\times$ & --- & 2 \\
2022-06-03 & \checkmark  & U & 3 \\
2022-06-15 & \checkmark  & U & 4 \\
2022-07-01 & \checkmark  & R & 5/C \\
2022-07-15 & \checkmark  & R & 6/D \\
2022-08-02  & $\times$ & --- & 7 \\
2022-08-14 & $\times$ & --- & 8 \\
2022-09-03 & $\times$ & --- & 9 \\
\hline
\multicolumn{4}{p{0.9\linewidth}}{ $^{*}$ L and R polarizations refer to left and right circular polarization, while U indicates unpolarized emission.}\\
\multicolumn{4}{p{0.9\linewidth}}{ $^{\dagger}$ This epoch showed a circularly polarized burst followed by an incoherent gyrosynchrotron flare.  }\\
\multicolumn{4}{p{0.9\linewidth}}{ $^{\ddagger}$ All detected polarized bursts last for at least 1~hour except this one, which lasts only $\sim$1~min.  }\\

\end{tabularx}
\end{center}
\end{table}

\subsection{Mapping Magnetic Footpoints}

\begin{figure*}
    \centering
    \includegraphics[width=0.8\textwidth]{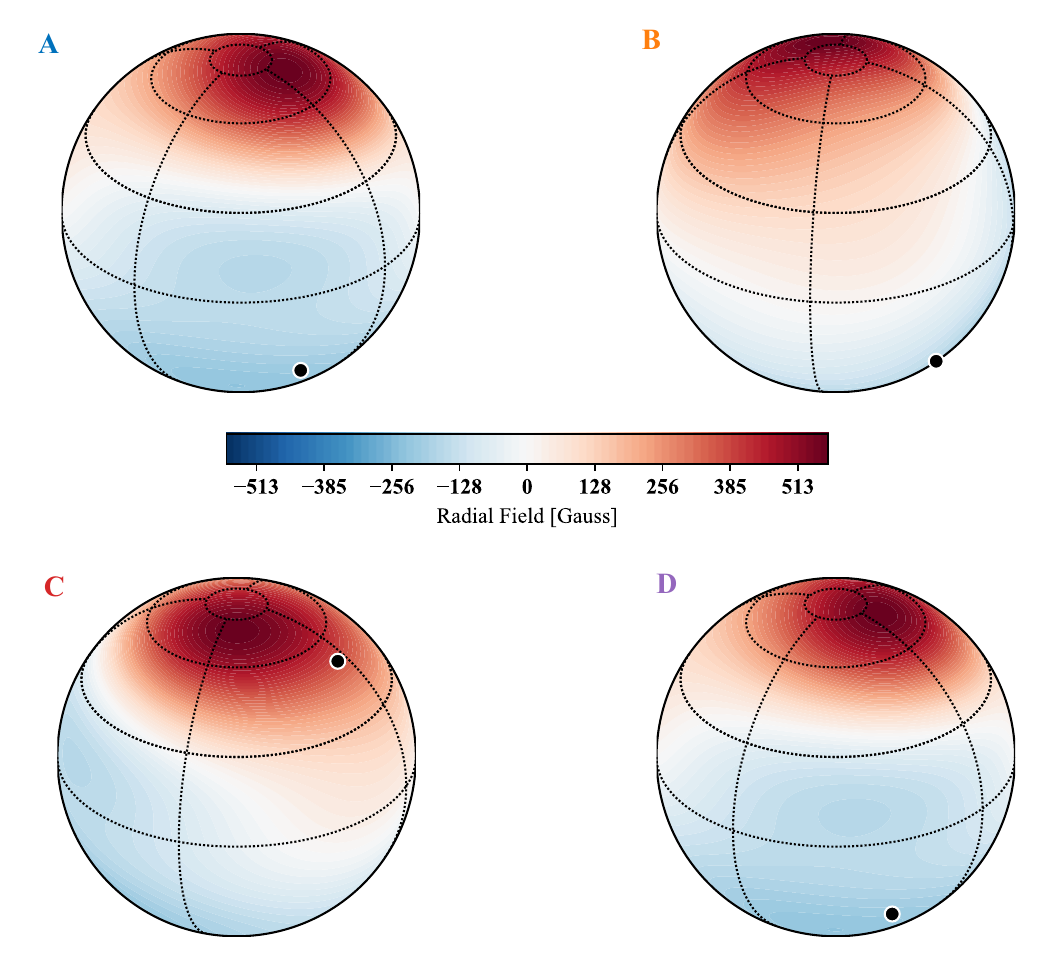}
    \caption{The stellar surface phased to the four dates corresponding to polarized radio burst detections (see Figure~\ref{fig:orbdiag}) shows the strength and polarization of the magnetic field at those epochs (red for positive field pointed away from the star). Each of the 4 panels shows a block dot indicative of the location of the magnetic footpoint mapped from the exoplanet position to the stellar surface. During all 4 epochs of polarized burst detections the exoplanet footpoint was visible to observers. }
    \label{fig:polfoot}
\end{figure*}

By phasing the star rotation and the planet orbit to the radio observation epochs, we can examine how YZ~Ceti~b connects magnetically to the stellar surface. Starting from the position of the planet at each radio epoch, we used the extrapolated magnetic field topology (see Section~\ref{sec:pfss}) to compute the 3D magnetic streamline tracing back to the host star. With the large-scale field dominated by the dipolar component, and the planet passing through open field lines of stellar wind, the footpoints generally map cleanly to magnetic polar regions on the stellar surface.

In Figure~\ref{fig:polfoot}, we show, for the four epochs with polarized radio detections, the projection of the stellar surface visible at that time, and the location of the footpoint mapping to the YZ~Ceti~b (black dots). The top two panels are VLA epochs and the bottom two are uGMRT epochs, with the color shading indicative of the radial magnetic field strength at the surface. For 3 of the 4 epochs, the footpoint is clearly in the visible hemisphere, whereas for epoch B, the footpoint attaches right at the limb. This unique footpoint mapping may explain how the epoch B burst lasts for only a couple minutes as compared to the hours long events in the rest of the polarized detection epochs. Semi-limited visibility of ECM sources could shorten the apparent duration of the event across a transient favorable beaming geometry. Interestingly, for epochs A and D, the planet is in a similar orbital position (see Figure~\ref{fig:orbdiag}), and the stellar face is at a similar rotational phase: repeated magnetic and orbital geometries produced repeated radio emission at the star-planet synodic period.

In contrast, the epochs without polarized emissions typically connect on the back side of the star, blocking the view to the magnetic footpoint, or may be disconnected from the star. The latter case occurs when the planet is positioned near the stellar wind current sheet and would be associated with passages across the Alfv\'{e}n surface (see Figure~\ref{fig:orborient}). During such instances, dependent on the relative phasing of both the planet orbit and stellar rotation, the streamlines do no connect back to the star. Our magnetic field extrapolations and the wind assumptions may only give an approximate answer, but our estimates are likely indicative of this real effect where the planet experiences significant changes in the interplanetary magnetic field as the orbit crosses the magnetic equator.

The qualitative summary of the magnetic footpoints at each radio epoch is shown in Table~\ref{tab:footp}, following the epoch labels shown in Table~\ref{tab:radio}. The conclusions apparent in this Table are also robust to the effects of near surface small scale fields that may shift the exact locations of the footpoints. We discuss the implications of our findings within the context of evaluating star-planet interactions for the YZ~Ceti system next.

\begin{deluxetable*}{l | ccccc | ccccccccc}
\centering
\tablecaption{Epoch Mappings \label{tab:footp}}
\tablehead{ \colhead{} &  \multicolumn{5}{c}{ \underline{ \citet{Pineda2023NatAs...7..569P} } } & 
            \multicolumn{9}{c}{\underline{ \hfill \citet{Trigilio2023arXiv230500809T} \hfill } } \\
            \colhead{ } & \colhead{ \#1 } &  \colhead{ \#2 } & \colhead{ \#3 } & \colhead{ \#4 } & \colhead{ \#5 } & \colhead{ \#1 } &  \colhead{ \#2 } &  \colhead{ \#3 } & \colhead{ \#4 } & \colhead{ \#5 } & \colhead{ \#6 } & \colhead{ \#7 } & \colhead{ \#8 } & \colhead{ \#9 } }
    \startdata
Detection\tablenotemark{a} & --- & R, U & U & --- & L &  --- & --- & U & U & R & R & ---& --- & --- \\
Footpoint\tablenotemark{b} & --- & F, F & B  & B & E & ---  & F & F & --- & F & F & B & F & B  \\
Current Sheet\tablenotemark{c}  & \checkmark & ---, $\sim$ & $\sim$ & --- & --- & \checkmark & --- & $\sim$ & \checkmark & --- & --- & --- & --- & ---  \\
Pole Connected\tablenotemark{d} & --- & S & N & S & S & --- & S & N & --- & N & S & N & N & S \\
North Pole Visibility\tablenotemark{e}  & Y & Y & Y & Y & N & Y  & Y & Y & N & Y & Y & N & N & Y \\
    \enddata
    \tablenotetext{a}{Same as polarization shown in Table~\ref{tab:radio}.}
    \tablenotetext{b}{Indicates whether the magnetic foot point was on the visible front of the star (F), the back (B) or near the terminator edge (E).}
    \tablenotetext{c}{Checkmark for current sheet proximity at epoch mid-point, tilda if epoch is crossing adjacent or duration encapsulates both connected and unconnected states, and empty otherwise.}
    \tablenotetext{d}{North/South depending on which magnetic polarity is connected to the planet position.}
    \tablenotetext{e}{Yes/No for whether the north magnetic pole is visibly pointed toward the observers or not. For example, at rotation phase 0 (see~Figure~\ref{fig:zdi}), the positive pole is tilted toward the observer further than the rotation axis (Y).}
\end{deluxetable*}

\section{Discussion} \label{sec:discuss}

Despite the suggestive recurrence of polarized radio emissions during epochs A and D, and the consistency with which observations at that same orbital phase have produced radio emissions, confirmation of magnetic star-planet interactions requires additional evidence and physical consistency across additional metrics. We evaluate those here. 

\subsection{Field Strengths and Observing Frequencies}

In Section~\ref{sec:spi}, we applied the new magnetic field measurements to compute SPI radio flux predictions, and compared to the measured radio burst densities. A self-consistent SPI scenario would have required planetary magnetic fields of at least several Gauss, with some leeway based on expectations for potentially underestimated stellar magnetic fields in ZDI. With the magnetic field topology, we can further examine the expected ECM frequency associated with the planet magnetic footpoints during the radio observations.  

 For epochs A, B and D, the footpoints map to the weaker southern pole (see Figure~\ref{fig:polfoot}), and give low ECM frequencies ($\lesssim 0.5$~GHz) in the low corona, for the assumption that the ZDI-detected field is not enhanced by small-scale field. Only for epochs  C and D with the uGMRT does the ECM fundamental reasonably map well to the proper radio observing band (without needing additional unaccounted-for magnetic field). If the southern pole matched the strengths seen in the visible northern magnetic pole, the footpoints associated with epochs A and B would still not match the observed radio frequencies.

Even though this mismatch between the fields implied by ZDI and observed radio data are not unprecedented, even when data are taken only months apart \citep[][]{Llama2018ApJ...854....7L}, potentially underestimated magnetic field strengths in ZDI may be insufficient to account for this discrepancy. The stellar magnetic field would have to have been a factor of 3 to 5 times stronger than is implied by the ZDI data. New measures of the small scale fields on YZ~Ceti (e.g.,with Zeeman broadening technique) would help evaluate the extent of this issue. ECM emissions occurring at low-level harmonics of the fundamental are also disfavored as too inefficient to produce the strong detected bursts.

Given the several year separation between the radio epochs and the spectropolarimetric data set, it is possible that the stellar magnetic field has evolved significantly. If the VLA radio bursts are caused by SPI, reaching a GHz ECM frequency requires either unaccounted-for small-scale magnetic field or a declining evolution in the magnetic field strength of YZ~Ceti between the VLA epochs and the ZDI data set. The uGMRT data and the SPIRou data are separated closer in time ($\sim$1 yr), and the observed field strengths are more closely matched to the radio frequencies of those polarized bursts. Long-term ZDI observations of YZ~Ceti are needed to assess the extent of any potential evolution to verify whether the past VLA observations could have been consistent with SPI.

\subsection{ECM Beaming Geometry}

The magnetic field geometry also enables a test of whether we should have expected observable ECM emissions at a given radio epoch based on its anisotropic beaming. Within the Solar System, ECM cone opening angles are large (70-90$^\circ$), but in theory could be narrower in more energetic electron beams \citep{MelroeDulk1982ApJ...259..844M, Lynch2015ApJ...802..106L}. 

\begin{figure}
    \centering
    \includegraphics[width=0.5\textwidth]{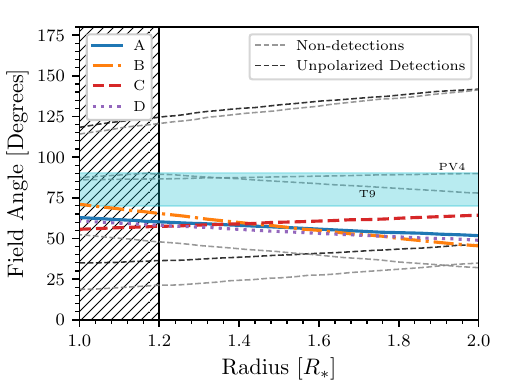}
    \caption{ The angle between the magnetic field direction along the flux tube and the observer, with the canonical ECM regime shaded in cyan. Angles above 90$^\circ$ are pointed away from the observer and out of view. The polarized detections are all cluster with footpoint angles in the range of 55-70$^\circ$. Not all radio epochs are shown here as those near the wind current sheet do not clearly map footpoints to the stellar surface (see Table~\ref{tab:footp}). The hatched region within 1.2 $R_{*}$ indicates where we may expect small scale magnetic fields to perturb baseline directions implied by the large-scale magnetic field.}
    \label{fig:beamgeo}
\end{figure}

In Figure~\ref{fig:beamgeo}, we show the angle between the magnetic field vector, connecting the star and planet, and our line of sight to the system. We include a hatched region to indicate where we may expect strong near surface small-scale magnetic field to disrupt the geometry of the large-scale field from ZDI. We approximated the maximum radial extent of the hatched region based on where magnetic field strengths of order $Bf = 2.2$~kG in spherical harmonics modes comparable to the smallest in the ZDI map ($\ell =10$), fall below the strength of the dipole component.

Observable ECM sources with solar system-like electron beams would cluster in the cyan shaded region. Instead, all of the polarized radio detections exhibit line-of-sight angles at the foot points in the range of 55-70$^\circ$. The non-detection epochs show a broad range of angles, with several footpoints either almost aligned with the line of sight (too narrow to see ECM) or pointed away from the observer. 

Interestingly, epochs PV4 and T9, which occur near the quadrature orbital position, have field angles near perpendicular. If SPI is consistently taking place for this system, then those epochs should have been viable times to observe planet-induced ECM emission. However, for these two epochs the footpoint is behind the stellar limb. Above the surface, the magnetic fields of these flux tubes correspond to ECM frequencies below the observing bands.

Alternatively, the observability of ECM and the combined non-detections and polarized detections would be consistent with SPI if the induced electron energy spectrum had energies of 30-110 keV, higher than what has been historically attributed to Solar System ECM sources \citep[$\sim$10~keV for Earth's AKR or Jupiter's HOM; e.g.,][]{Wu1979ApJ...230..621W, Treumann2006A&ARv..13..229T, Louarn2017GeoRL..44.4439L}. Stellar ECMI sources have now been detected with energies of 20-30 keV \citep{Zarka2025A&A...695A..95Z}. And, even the Juno mission has discovered narrower radio beaming angles from higher energy electron populations ($\sim$50~keV) within the Jupiter-Io flux tube \citep{Martos2020JGRE..12506415M}. At these higher energies, the effective ECM beaming geometry would shift the cyan region of Figure~\ref{fig:beamgeo} down to 55-70$\degr$. The consistent clustering of the polarized radio detection angles would further imply that the SPI is consistently generating electrons of 30-110 keV energies across the different epochs. This clustering persists up to higher altitudes, making it robust to the influence of missing small scale field. Thus, in Figure~\ref{fig:beamgeo}, we present the typical angles associated with potential SPI, and a baseline from which to examine perturbations associated with future measurements of the (small-scale) near surface field.

With narrower beaming angles for the ECM emission cone, the total solid angle into which this light is emitting gets reduced. If true, this directly impacts the solid angle ($\Omega$) assumption used in Equation~\ref{eq:rflux}, allowing for an increase in the theorized SPI flux of $\sim$20\%. Although this does not fully address the tension in the SPI scenario for implied exoplanetary magnetic fields (see Section~\ref{sec:spi}), it would move the predictions toward greater consistency by shifting up the curves in Figure~\ref{fig:predbplanet}.

Near quadrature, the line of sight angle would be too big to view ECM sources, but just right for the footpoints at epochs A, B, C, and D. If typical electron beams outside the Solar System are stronger than anticipated, observations just at orbital quadrature might miss SPI producing systems in SPI focused surveys. A better understanding of the expected electron spectrum along the connecting flux tube generated by a close-in exoplanet with its host star would help evaluate the plausibility of such conditions. 

\subsection{Radio Epochs and Polarization}

In SPI, the polarization properties of the radio detections should be defined by the surface magnetic field polarity. For planet-induced ECM emission, we would expect right-hand circularly polarized light (x-mode) coming from the north (positive) magnetic pole of the star \citep{Melrose1984JGR....89..897M,Mutel2007JGRA..112.7211M}, and left-handed polarization from the south.\footnote{Strong field strengths and free-free absorption likely rule out plasma emission scenarios \citep{Pineda2023NatAs...7..569P}.} Examining the footpoints for the polarized detection epochs, we see that only epochs B and C are consistent with this paradigm. 

Epochs A and D, which show the clearest synodic periodicity exhibit footpoint connections to the southern magnetic pole, and yet were seen as right-handed circularly polarized bursts. The polarization properties of the bursts in epochs A, C, and D, could be explained if the bursts all came from stochastic stellar processes in events localized in the regions of positive polarity that show the strongest visible radial magnetic field strengths. Unlike epoch B, the north magnetic pole of the star was clearly visible (see Table~\ref{tab:footp}) during each of those radio observations (see Figure~\ref{fig:polfoot}). Event B would then have to be associated with regions of negative radial field and weaker strengths, yet still capable of generating a strong burst albeit of shorter duration. While the event rate of such stochastic stellar bursts remains unknown, more radio detections are needed to distinguish between these possibilities, and discern SPI false-positives.

Since we assumed a magnetic field extrapolation as a PFSS, with open field lines beyond 5$R_{*}$, the planet footpoint connects to only one stellar hemisphere when possible. While less likely at an orbit of $\sim$21$R_{*}$, if the field was actually closed to these distances, the exoplanet could be connected to both hemispheres assuaging these polarization concerns. However, under the SPI hypothesis we suggest that it is more likely that evolution in the magnetic field topology may explain the inconsistencies in the polarization data.

\begin{figure*}
    \centering
    \includegraphics[width=0.6\textwidth]{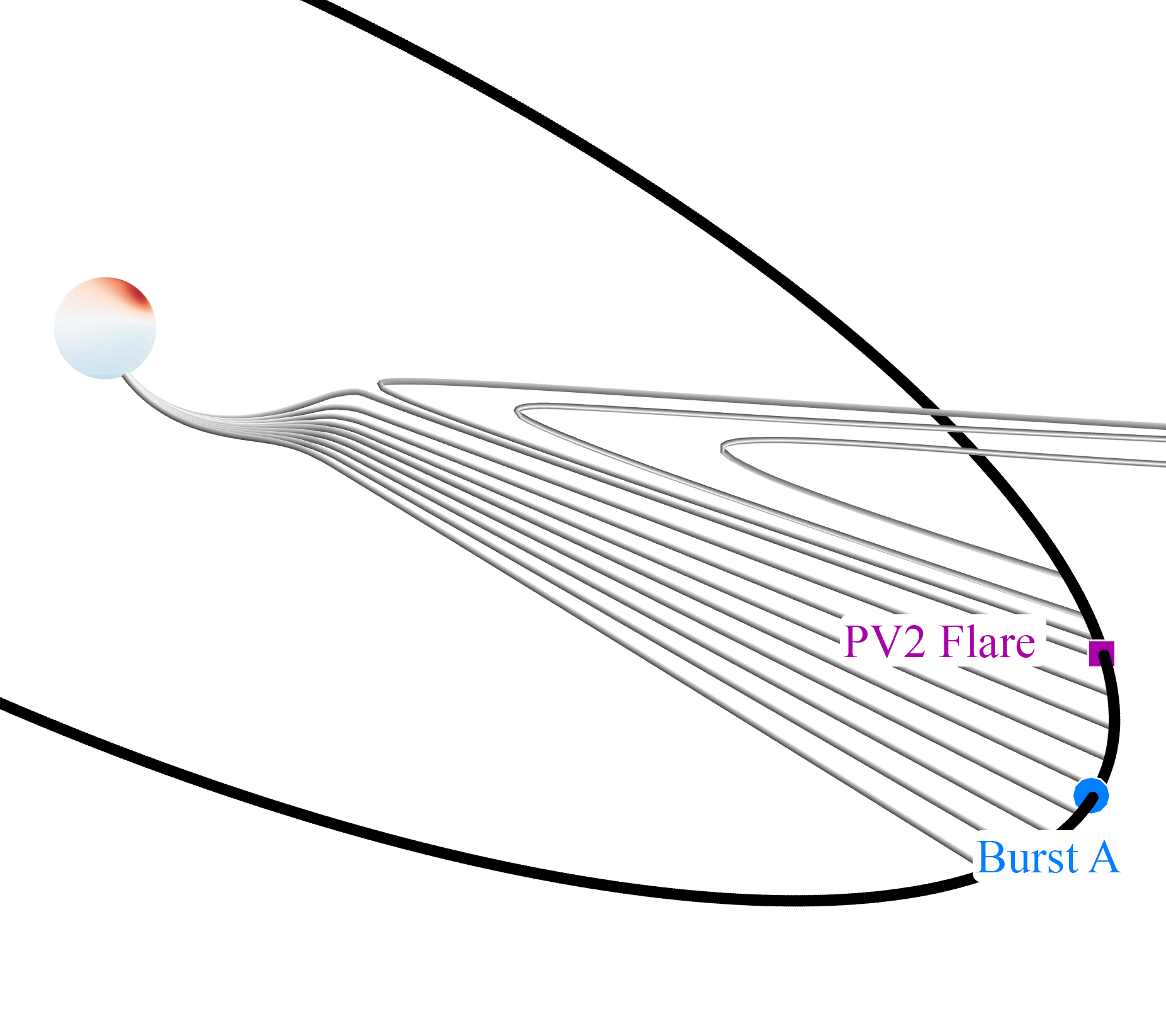}
    \caption{Mapping the magnetic streamlines (gray) between YZ~Ceti~b (orbit in black) and its host star during the monitoring of radio epoch PV2 (see Table~\ref{tab:radio}). At the start of the observation, the footpoints clearly map to the southern magnetic pole, but by the end of the observation the planet has moved close enough to the current sheet (across the Alfv\'{e}n surface) that the connectivity ceases. The blue circle shows the planet position during burst A, while the square denotes the planet position near the start of the flare during epoch PV2 \citep[see within][]{Pineda2023NatAs...7..569P}. We suggest that when the planet approaches the current sheet, these changes in the magnetic connectivity could manifest as planet-induced flares (see Section~\ref{sec:plflare}).}
    \label{fig:flarespi}
\end{figure*}

\subsection{Planet-Induced Flaring}\label{sec:plflare}

Recent studies have also raised the possibility that close-in satellites could be inducing flares on their host stars \citep{Loyd2023AJ....165..146L, Fitzmaurice2024AJ....168..140F, Ilin2024MNRAS.527.3395I,Ilin2025arXiv250700791I}. The mechanism is likely similar to that studied in periodically enhanced chromospheric activity \citep[e.g.,][]{Shkolnik2008ApJ...676..628S, Cauley2019NatAs...3.1128C}, but now based on elevated flare statistics. This observational signature could emerge as a consequence of extra magnetic energy being deposited into the stellar upper atmosphere serving to trigger reconnection. The flares in the radio would present as frequent unpolarized gyrosynchrotron bursts. This kind of SPI is distinct from the auroral analog ECM bursts (events A, B, C, D) discussed in this paper and in similar works searching for SPI \citep[e.g.,][]{PerezTorres2021A&A...645A..77P}. Nevertheless, this idea is worth discussing here in the context of YZ~Ceti and its magnetic field topology, as this host does appear to exhibit an elevated flare rate given its rotation period and mass \citep{Fitzmaurice2024AJ....168..140F}.

The cumulative radio data sets for YZ~Ceti include epochs of unpolarized emissions as seen by both \citet{Pineda2023NatAs...7..569P} and \citet{Trigilio2023arXiv230500809T}. These authors observed 4 unpolarized events across the 14 epochs of radio monitoring. While these events are insufficient for a clear statistical test of radio flaring, we can look for insights into the SPI hypothesis with comparison to the magnetic field topology.

One of the epochs of unpolarized emissions, T4, appears to be consistent with proximity to the current sheet, where the exoplanet approaches the magnetic equator and the radial field of the stellar wind switches direction. This orbital location is likely associated with crossings of the Alfv\'{e}n surface (see Figure~\ref{fig:orborient}). When we examined each of the other unpolarized epochs, we noticed a similar phasing of the orbital position and stellar magnetic field. The T3 detection takes place just after the conclusion of an equatorial crossing. During the longer PV3 epoch, YZ~Ceti~b is connected to the northern pole but moves closer towards the Alfv\'{e}n surface by the end of the observing window. During that epoch, the radio emissions are persistently variable but generally stronger at the end of the epoch \citep{Pineda2023NatAs...7..569P}. In epoch PV2, after the polarized burst, there was a strong flare captured from its onset which appears to extend beyond the end of the observations. During this epoch YZ~Ceti~b begins connected to the southern magnetic pole, and by its conclusion also approaches a current sheet crossing. 

In Figure~\ref{fig:flarespi}, we show the system magnetic geometry during epoch PV2, with the times associated with the polarized burst and flare onset indicated with a blue circle and purple square, respectively. The magnetic streamlines clearly map to the star for the majority of that epoch, but disconnect by the end of the observation. Because there is some uncertainty, in both the orbital geometry, its ephemeris, and the stellar magnetic field evolution between epochs, the phasing results here cannot be treated as exact, but are rather suggestive of how SPI could manifest in this system.

Besides the unpolarized radio detection epochs, there were two additional observations (T1 and PV1) which were indicative of current sheet proximity but did not show detected emissions. For times around the T1 epoch, the most proximal footpoints appear to be on the back side of the star, making triggered flares unobservable. Epoch PV1 likely suffered from the same issue despite corresponding to very similar orbital and stellar phases as the unpolarized detection seen in PV3; the time separation between epochs is a few days (see Table~\ref{tab:radio}). The remaining radio non-detections are consistent with this picture, with orbital positions away from the current sheet in clearly sub-Alfv\'{e}nic field displaying no emission even when having visible footpoints on the stellar surface (e.g., T2, T8).

By crossing the Alfv\'{e}n surface near the magnetic equator, the exoplanet would experience a significant change in the connectivity to the host star, and this could then trigger the energy release necessary to induce flares on the host star, either entering or exiting the sub-Alfv\'{e}nic regime. In principle, these events may be initiated with reconnection near the planet itself, but given the GHz frequencies of the observed unpolarized emission and their total energy, we expect the flaring to correspond to source regions near the stellar surface. Such triggers at the planet location may inject plasma back into the flux tube with electrons spiraling toward that star that could then generate detectable radio emissions, analogous to some interacting binary stars \citep{Massi2008A&A...480..489M}. If there is a regular cadence to reconnection with the exoplanetary magnetic field, then the Alfv\'{e}n surface crossings may also initiate strong radio aurorae in the exoplanet magnetic field itself at the system synodic period. Such a possibility could make YZ~Ceti a promising target for low-frequency searches for exoplanet aurorae \citep[e.g.,][]{Zarka2012sf2a.conf..687Z, Burns2019arXiv191108649B, Turner2021A&A...645A..59T}.

How much energy could be released by planet-induced flaring and its dependence on planet and stellar magnetic properties is unclear, but would define a new case of SPI in this system. 
Additional radio monitoring is essential to build up the statistics of such events, but it is clear that this SPI 
should have a strong dependence on the synodic period of the system and the geometry of the magnetic field.

Results showing enhanced flaring activity in optical data sets for slowly rotating M dwarf stars \citep[e.g.,][]{Fitzmaurice2024AJ....168..140F} could be explained through the recurrence of satellite current sheet crossings as suggested here for YZ~Ceti. With potentially variable strengths, every crossing may not have generated detectable optical flares (e.g., TESS), but the mechanism should enhance the overall observed flare rate. In such systems, the observability and periodicity is a complicated function of the magnetic field and the orbital geometry, and will require extended monitoring with concurrent magnetic field topology measurements to discern clearly.

\subsection{Evaluating SPI}

Given the discussions above, what conclusions can we draw regarding magnetic star-planet interactions in the YZ~Ceti system? We summarize the evidence for and against the SPI scenario below.

\vspace{0.5cm}

\emph{Evidence in support of SPI:}

\begin{itemize}
    \item Sub-alfv\'{e}nic SPI models can explain radio flux densities to within factors of a few
    \item Orbital phasing of polarized radio bursts (Epochs A, B, C, D)
    \item Synodic phasing of polarized radio bursts (Epochs A, D)
    \item Radio emission at ECM fundamental frequency (C, D)
    \item SPI scenario consistent across detections and non-detections for electron beams of $\sim$30-110~keV
    \item Unpolarized flares show tentative association with orbital position and current sheet crossings 
\end{itemize}

\emph{Evidence against SPI:}

\begin{itemize}
    \item Radio emission is not at ECM fundamental frequency (Epochs A, B)
    \item ECM beaming geometry inconsistent with sources emitting into cones with solar system-like (near 90$\degr$) opening angles
    \item 50\% of polarized burst detections show polarizations inconsistent with magnetic footpoints
    \item Required exoplanetary field strengths are high compared to expectations at expected planet mass and radius for YZ~Ceti~b
\end{itemize}

Together, these findings present a mixed picture of SPI from YZ~Ceti. While the phasing and geometry appears consistent with the SPI hypothesis, the ECM properties associated with planet-induced bursts do not completely match the properties of the observations. However, measurements of the large-scale magnetic field with ZDI could have ruled out SPI altogether, if the field was especially weak (ruling out a sub-Alfv\'{e}nic system), or if the polarized epochs clearly mapped to footpoints behind the star. Instead, the data illustrate complexities to the system, but clearly suggest that SPI is plausible. 

Moreover, by comparing the orbital timing of unpolarized radio flare-like emissions with the system magnetic topology, we suggest that this star may be exhibiting planet-induced flaring (see Section~\ref{sec:plflare}). While ECM radio emission should be persistent (even if variable), with observability defined by the anisotropic beaming, planet-induced flaring is isotropic and thus could be present for specific combinations of orbital and stellar phases. Namely, those positions traced here as associated with current sheet crossings. Each mechanism is distinct, but both would depend on the synodic phasing of the system. 

SPI confirmation in either case requires additional detections of radio bursts (or general flaring), to provide a strong statistical argument, but the strong impact of the magnetic field on the observability suggests the stellar rotational phase should be taken into account when planning and interpreting such observations. Crucially, concurrent spectropolarimetry data will be essential to help evaluate the nature of potential SPI signatures. 

The importance of contemporaneous radio and ZDI data sets is underscored by clear evidence for magnetic field evolution in stars with similar fundamental properties to YZ~Ceti \citep{Lehmann2024}. While our analysis assumes relative stability of the magnetic field topology over several years, there likely has been some evolution of the stellar field. Indeed, a consistent SPI scenario of polarized burst properties requires significant evolution of the overall magnetic field strength, which would also imply more plausible magnetic field strengths for YZ~Ceti~b (see Section~\ref{sec:spi}).

Nevertheless, possible shifts in the topology and strength across epochs introduces additional uncertainty into the geometric/topological analysis presented in this work. This may be compounded by unknowns in the 3D orientation of the planet orbit, although we have assumed a typical scenario with aligned orbital and stellar rotational axes. For this RV detected system, such unknowns are inherent to the data, although updated ephemerides would help evaluate the timing of the bursts with the topology. Future detections of SPI behavior in transiting systems would alleviate these concerns. For the SPI hypothesis, changes in the magnetic topology may help explain some of the discrepancies in the polarized burst behavior from YZ~Ceti, but will increase uncertainty in mapping topology across epochs. In the case of planet-induced flaring, if it is associated with exoplanet crossings of the stellar wind current sheet, the observational signatures and timing will be tied to the location of the magnetic equator. If this location is more persistent across stellar magnetic field evolution, planet-induced flaring may still present a clearly periodic signature of SPI robust to unknown field evolution.

These concerns are difficult to address without contemporaneous ZDI maps to match to radio observation epochs. Long-term monitoring will further help evaluate the extent and nature of magnetic field changes, and help to corroborate statistical confirmation of orbitally modulated stellar bursts. Accounting for that temporal variability of the stellar magnetic field will then reinforce any conclusions regarding SPI. Regardless, our analysis demonstrates how the ZDI topology can inform SPI considerations, and the various checks required when evaluating polarized radio bursts as signatures of star-planet interaction. For YZ~Ceti, the SPI case remains unconfirmed.  Nevertheless, our study suggests that the topology does not rule out SPI, and moreover the measurements imply that SPI could be plausible and even favorable depending on the nature and extent of the host magnetic field evolution across several years.

\section{Summary} \label{sec:summary}

With this investigation, we have measured the large-scale magnetic field of exoplanet host YZ~Ceti through Zeeman Doppler Imaging (Section~\ref{sec:zdi}), revealing a predominately dipolar field of $\sim$500~G at the peak northern pole with a surface averaged strength of $\sim$200~G. We further evaluated how its topology informs the ability of the system to exhibit signatures of magnetic star-planet interactions (Section~\ref{sec:geom}). Confirmed SPI from any system would provide a powerful tool to examine stellar host and exoplanet magnetic properties (Section~\ref{sec:spi}), and as such we assessed whether the measured field topology supports or refutes the case of SPI from the YZ~Ceti system (Section~\ref{sec:discuss}). By considering both detection and non-detection epochs, we provided a comprehensive consideration of SPI radio bursts in the context of the host magnetic field topology.

The phasing of the orbital period and stellar rotation with repeated radio bursts favors the SPI scenario, independent of the exact magnetic topology. However, to be fully consistent with SPI, the system requires both stronger electron beams than are typically seen in the Solar System, and significant evolution in the magnetic field strength over several year timescales without dramatic shifts in the star-planet connectivity. Long-term magnetic monitoring of YZ~Ceti will be required to assess this prospect. 

Our examination of the radio detections and the topology also revealed that several of the unpolarized radio detections could be associated with planet orbital positions near crossings of the magnetic equator and likely the Alfv\'{e}n surface. Such crossings could release magnetic energy and trigger flare-like events on the host star. The uncertainty in the topology across epochs makes this idea merely suggestive. However, we hypothesize that this system could be exhibiting planet-induced flares \citep[see also][]{Fitzmaurice2024AJ....168..140F}, with Alfv\'{e}n surface crossings defining the means of magnetic energy release that potentially triggers reconnection in the stellar atmosphere.

Despite the mixed results, with this article we have illustrated the utility of magnetic field topology measurements in examining star-planet interactions, and demonstrated the various metrics that must be tested when evaluating the possibility of magnetic SPI as seen with radio data sets. Additional monitoring of YZ~Ceti is required to bolster the case for SPI, but it remains an exemplary case study for understanding and testing star-planet interaction scenarios.

\begin{acknowledgments}

We thank the anonymous referee for useful discussions in the preparation of this article. This material is based upon work supported by the National Science Foundation under Grant No.\ AST-2108985 (JSP) and AST-2150703 (JRV). The National Radio Astronomy Observatory is a facility of the National Science Foundation operated under cooperative agreement by Associated Universities, Inc. AAV and SB acknowledge funding by the Dutch Research Council (NWO) under the project ``Exo-space weather and contemporaneous signatures of star-planet interactions" (with project number OCENW.M.22.215 of the research programme ``Open Competition Domain Science- M"). AAV acknowledges funding from the Dutch Research Council (NWO), with project number VI.C.232.041 of the Talent Programme Vici and from the European Research Council (ERC) under the European Union's Horizon 2020 research and innovation programme (grant agreement No 817540, ASTROFLOW). CPF acknowledges funding from the European Union's Horizon Europe research and innovation programme under grant agreement No. 101079231 (EXOHOST), and from the United Kingdom Research and Innovation (UKRI) Horizon Europe Guarantee Scheme (grant number 10051045).

Based on observations obtained at the Canada-France-Hawaii Telescope (CFHT) which is operated by the National Research Council of Canada, the Institut National des Sciences de l'Univers of the Centre National de la Recherche Scientique of France, and the University of Hawaii. The observations at the CFHT were performed with care and respect from the summit of Maunakea which is a significant cultural and historic site. 

This work has made use of the VALD database, operated at Uppsala University, the Institute of Astronomy RAS in Moscow, and the University of Vienna

\end{acknowledgments}

\software{ astropy \citep{Astropy2013A&A...558A..33A,Astropy2018AJ....156..123A,Astropy2022ApJ...935..167A}, specpolFlow \citep{Folsom2025}, tecplot
          }


\appendix

\section{YZ~Ceti as a multi-planet System}\label{sec:ap}

Throughout the main article we considered the case for star-planet interactions in the YZ~Ceti system with the focus on the innermost planet as the prime candidate. For completeness, we also consider here whether the outer two planets in the system are capable of powering SPI. As discussed in \citet{Pineda2023NatAs...7..569P}, for the weak wind case, the full multi-planet system may be under sub-Alfv\'{e}nic conditions.

\begin{figure*}
    \centering
    \includegraphics[width=0.48\textwidth]{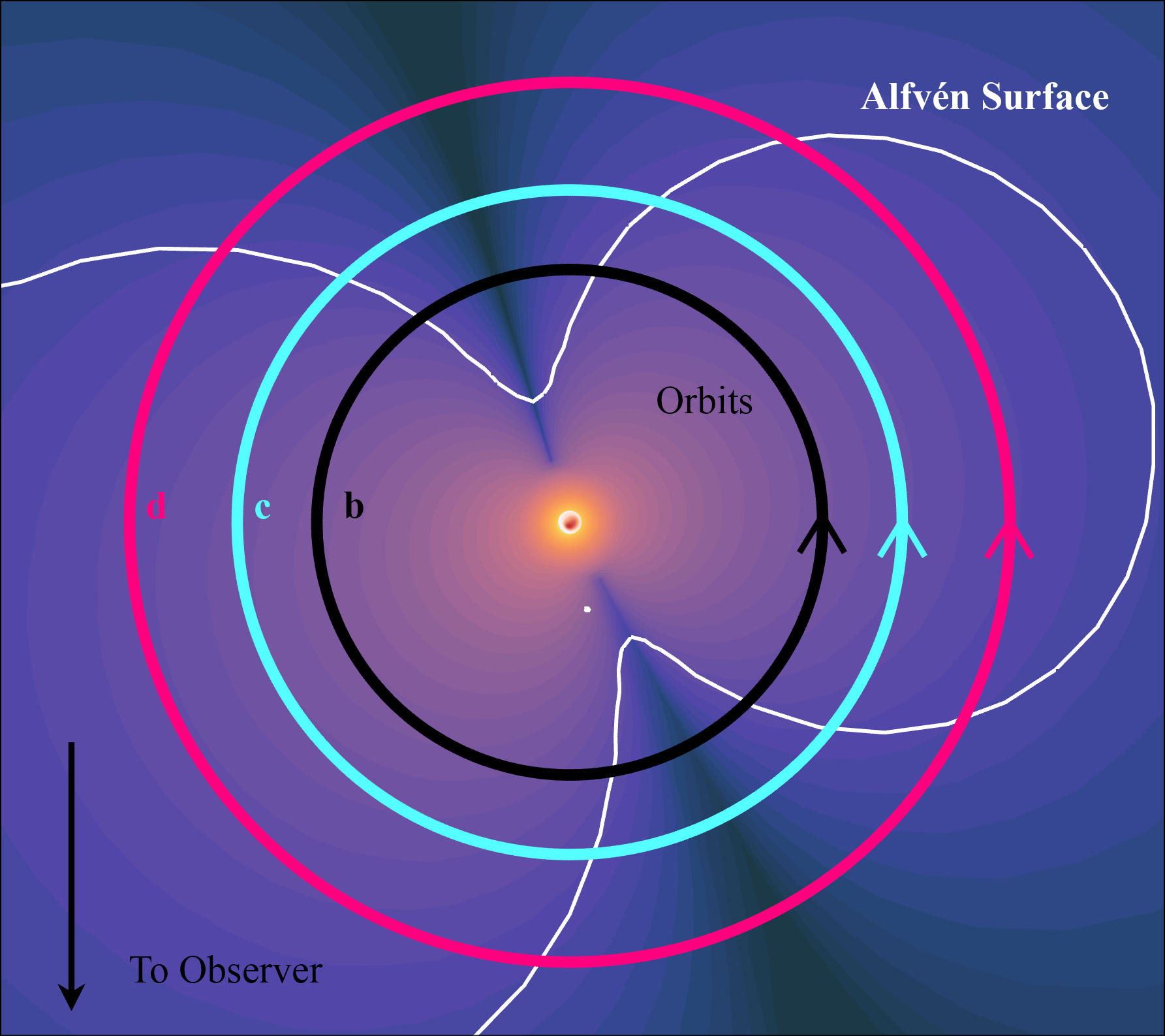}
    \includegraphics[width=0.48\textwidth]{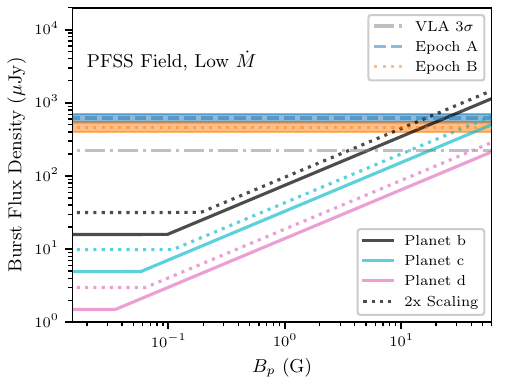}
    \caption{\textit{Left} - Same as the right panel of Figure~\ref{fig:orborient}, but now also showing the orbits of the two outer planets. The asymmetry in the Alfv\'{e}n surface is a consequence of the tilted magnetic field. YZ~ceti~c and d spend more time in super Alfv\'{e}nic regimes than planet b. \textit{Right} - Same as Figure~\ref{fig:predbplanet} for the predicted radio flux densities as a function of intrinsic planetary magnetic field, now including the cases for the outer two planets. The dotted lines also show the potential strength of the radio bursts for the same 2x scaling utilized in Section~\ref{sec:scaleBspi}. }
    \label{fig:ap_multi}
\end{figure*}

In the left panel of Figure~\ref{fig:ap_multi}, we show the stellar Alfv\'{e}n surface of the system based on the ZDI reconstruction, similar to Figure~\ref{fig:orbdiag} with the locations of all three planets in the system. All satellites experience sub-Alfv\'{e}nic conditions and may also cross the Alfv\'{e}n surface. Because the more distant planets are farther away, they experience weaker stellar magnetic fields, and less dense winds, which generally limit the strength of their potential SPI signatures. 

This is evident in the right panel of Figure~\ref{fig:ap_multi}, where we show the strength of potential polarized radio bursts as discussed in Section~\ref{sec:spi}. The solid curves are defined for each exoplanet with radii consistent with Earth density at the measured minimum mass and the set inclination (60$^{\circ}$). YZ~Ceti~c and d might produce bursts that would be at least factors of several weaker than those induced by YZ~Ceti~b. Planetary field strengths of $>$50~G would be required to generate the polarized burst detections seen by the VLA. Uncertainties in the stellar wind properties and similar assumptions would scale these curves up and down together. However, in multi-planet systems under sub-Alfv\'{e}nic conditions, wing-wing interactions could enhance the overall effects relative to what each planet can do individually \citep{Fischer2019ApJ...872..113F,Fischer2022A&A...668A..10F}. 

If Alfv\'{e}n surface crossings by planets c and/or d contribute to perturbations associated with planet-induced flaring (see Section~\ref{sec:discuss}), the strength of such perturbations are similarly weaker and less frequent, as compared to the effects of YZ~Ceti~b. New theoretical treatments of the consequences of the crossings could help determine whether c and d may be implicated with SPI. Nevertheless, the radio signatures potentially associated with SPI from planets c and d will require more sensitive instruments and additional monitoring to discern clearly. If detected in future, a multi-planet nature to the SPI would serve as the perfect probe of stellar wind physics, with multiple markers of the wind properties as it accelerates through the M-dwarf system.

\section{Data Log}

\begin{table*}[ht]
\begin{center}
\caption{SPIRou observations of YZ~Ceti. The columns are: (1 and 2) date and universal time of the observations, (3) heliocentric Julian date minus the date of the first observation, (4) rotational cycle of the observations phased at $P_\mathrm{rot}=68.4$\,d and $\mathrm{HJD}_0=2460244.830417$, (5) exposure time of a polarimetric sequence, (6) signal-to-noise ratio at 1650 nm per polarimetric sequence, (7) RMS noise level of Stokes $V$ signal in units of unpolarised continuum. \label{tab:spirou_log}}
\begin{tabular}{c r c c c c c} 
\hline 
Date & UT & HJD & $n_\mathrm{cyc}$ & $t_{exp}$ & S/N & $\sigma_\mathrm{LSD}$\\
 & [hh:mm:ss] & & & [s] & & [$10^{-4}I_c$] \\
\hline
October 27 2023 & 7:54:24.04 & 0.0000 & 0.00 & 4x150 & 284 & 16.2\\
October 28 2023 & 8:33:15.82 & 1.0271 & 0.01 & 4x150 & 286 & 16.5\\
October 29 2023 & 9:49:28.45 & 2.0800 & 0.03 & 4x150 & 275 & 15.1\\
October 30 2023 & 11:34:03.87 & 3.1526 & 0.05 & 4x150 & 275 & 15.8\\
October 31 2023 & 8:54:50.20 & 4.0421 & 0.06 & 4x150 & 263 & 14.5\\
November 01 2023 & 9:26:50.99 & 5.0643 & 0.07 & 4x150 & 275 & 15.7\\
November 02 2023 & 9:17:26.56 & 6.0577 & 0.09 & 4x150 & 288 & 15.8\\
November 03 2023 & 9:56:38.02 & 7.0850 & 0.10 & 4x150 & 288 & 16.8\\
November 04 2023 & 7:44:18.59 & 7.9931 & 0.12 & 4x150 & 295 & 16.4\\
November 05 2023 & 7:14:14.36 & 8.9722 & 0.13 & 4x150 & 283 & 17.5\\
November 06 2023 & 9:15:13.39 & 10.0562 & 0.15 & 4x150 & 229 & 19.0\\
November 07 2023 & 7:20:25.74 & 10.9765 & 0.16 & 4x150 & 187 & 22.2\\
November 08 2023 & 9:38:42.99 & 12.0725 & 0.18 & 4x150 & 254 & 16.5\\
November 17 2023 & 6:30:48.21 & 20.9420 & 0.31 & 4x150 & 279 & 13.4\\
November 18 2023 & 7:40:46.90 & 21.9906 & 0.32 & 4x150 & 247 & 13.4\\
November 20 2023 & 7:44:35.98 & 23.9933 & 0.35 & 4x150 & 143 & 24.4\\
November 20 2023 & 8:04:40.91 & 24.0072 & 0.35 & 4x150 & 148 & 19.6\\
November 21 2023 & 7:10:32.75 & 24.9696 & 0.36 & 4x150 & 210 & 16.3\\
November 22 2023 & 5:59:14.08 & 25.9201 & 0.38 & 4x150 & 265 & 17.9\\
November 23 2023 & 5:41:02.69 & 26.9075 & 0.39 & 4x150 & 249 & 19.2\\
November 24 2023 & 5:01:13.60 & 27.8799 & 0.41 & 4x150 & 279 & 19.0\\
November 25 2023 & 5:31:49.01 & 28.9010 & 0.42 & 4x150 & 288 & 20.8\\
November 26 2023 & 4:39:34.24 & 29.8648 & 0.44 & 4x150 & 279 & 20.5\\
December 05 2023 & 5:21:35.44 & 38.8939 & 0.57 & 4x150 & 290 & 17.5\\
December 06 2023 & 4:55:53.07 & 39.8761 & 0.58 & 4x150 & 271 & 18.1\\
December 07 2023 & 4:24:00.40 & 40.8540 & 0.60 & 4x150 & 287 & 17.5\\
December 08 2023 & 4:25:01.97 & 41.8547 & 0.61 & 4x150 & 213 & 17.7\\
December 10 2023 & 7:30:43.02 & 43.9836 & 0.64 & 4x150 & 259 & 14.8\\
December 11 2023 & 6:59:54.55 & 44.9622 & 0.66 & 4x150 & 290 & 16.4\\
December 12 2023 & 5:53:20.50 & 45.9160 & 0.67 & 4x150 & 284 & 15.8\\
December 13 2023 & 7:23:47.53 & 46.9788 & 0.69 & 4x150 & 254 & 14.9\\
December 14 2023 & 7:21:21.47 & 47.9772 & 0.70 & 4x150 & 265 & 16.3\\
December 15 2023 & 5:52:54.43 & 48.9157 & 0.71 & 4x150 & 280 & 18.8\\
December 16 2023 & 7:42:17.40 & 49.9916 & 0.73 & 4x150 & 288 & 18.7\\
December 17 2023 & 5:34:17.34 & 50.9027 & 0.74 & 4x150 & 280 & 19.8\\
December 24 2023 & 4:48:06.14 & 57.8707 & 0.84 & 4x150 & 258 & 22.5\\
December 25 2023 & 4:49:02.65 & 58.8714 & 0.86 & 4x150 & 273 & 23.6\\
December 26 2023 & 4:33:16.39 & 59.8603 & 0.87 & 4x150 & 293 & 21.5\\
December 27 2023 & 4:33:14.70 & 60.8604 & 0.89 & 4x150 & 249 & 20.2\\
December 28 2023 & 4:34:18.50 & 61.8611 & 0.90 & 4x150 & 288 & 17.2\\
December 29 2023 & 4:34:00.47 & 62.8609 & 0.92 & 4x150 & 262 & 17.7\\
January 24 2024 & 4:44:04.80 & 88.8679 & 1.30 & 4x150 & 291 & 11.6\\
January 25 2024 & 5:09:45.87 & 89.8857 & 1.31 & 4x150 & 194 & 15.2\\
\hline 
\end{tabular}
\end{center}
\end{table*}

\bibliography{main}
\bibliographystyle{aasjournal}

\end{document}